\documentclass[a4paper,fleqn,usenatbib]{mnras}

\usepackage{newtxtext,newtxmath}

\usepackage[T1]{fontenc}
\usepackage{ae,aecompl}

\usepackage{longtable}

\usepackage{balance}

\usepackage{graphicx}	
\usepackage{amsmath}	
\usepackage{amssymb}	

\newcommand{\teff}{$T_{\rm eff}$}
\newcommand{\logg}{log\,{\it g$_\star$}}
\newcommand{\cai}{Ca\,{\sc I}}
\newcommand{\mgi}{Mg\,{\sc I}}
\newcommand{\vsini}{$v$\,sin\,$i$}

\usepackage{enumerate}
\usepackage[british]{babel}

\title[]{HD~89345: a bright oscillating star hosting a transiting warm Saturn-sized planet observed by K2}
\vspace{-0.5cm}
\author[V. Van Eylen et al.]{V. Van Eylen$^{1}$\thanks{E-mail: vaneylen@strw.leidenuniv.nl},
F.\,Dai$^{2,3}$,
S.\,Mathur$^{4,5}$,
  D.\,Gandolfi$^{6}$,
S.\,Albrecht$^{7}$,
M.\,Fridlund$^{8,1}$,
\newauthor	
R.\,A.\,Garc\'\i a$^{9,10}$,
E.\,Guenther$^{11}$,
M.\,Hjorth$^{7}$,
A.\,B.\,Justesen$^{7}$,
J.\,Livingston$^{12}$,
M.\,N.\,Lund$^{7}$,\newauthor
F.\,P\'erez Hern\'andez$^{4,5}$,
J.\,Prieto-Arranz$^{4,5}$,
C.\,Regulo$^{4,5}$,
L.\,Bugnet$^{9,10}$,
M.\,E.\,Everett$^{13}$,\newauthor
T.\,Hirano$^{14}$,
D.\,Nespral$^{4,5}$,
G.\,Nowak$^{4,5}$,
E.\,Palle$^{4,5}$,
V.\,Silva Aguirre$^{7}$,
T.\,Trifonov$^{15}$,\newauthor
J.\,N.\,Winn$^{3}$,
O. Barrag\'an$^{6}$, 
P.~G. Beck$^{4,5}$, 
W. J. Chaplin$^{16,7}$, 
W. D. Cochran$^{17}$,\newauthor
S. Csizmadia$^{18}$, 
H. Deeg$^{4,5}$,
M. Endl$^{17}$, 
P. Heeren$^{22}$,
S. Grziwa$^{19}$,  
A.\,P.\,Hatzes$^{11}$,\newauthor
D.\,Hidalgo$^{4,5}$, 
J.\,Korth$^{19}$, 
S.\,Mathis$^{9,10}$,
P.\,Monta\~nes\,Rodriguez$^{4,5}$, 
N. Narita$^{12,20,21}$, \newauthor
M.\,Patzold$^{19}$, 
C.\,M.\,Persson$^{8}$, 
F. Rodler$^{23}$,
A.\,M.\,S.\,Smith$^{18}$
\\
$^{1}$Leiden Observatory, Leiden University, postbus 9513, 2300RA Leiden, The Netherlands\\
$^{2}$Department of Physics and Kavli Institute for Astrophysics and Space Research, Massachusetts Institute of Technology,\\
Cambridge, MA, 02139, USA\\
$^{3}$Department of Astrophysical Sciences, Princeton University, 4 Ivy Lane, Princeton, NJ, 08544, USA\\
$^{4}$Departamento de Astrof\'isica, Universidad de La Laguna, E-38206, Tenerife, Spain\\
$^{5}$Instituto de Astrof\'isica de Canarias, C/ V\'ia L\'actea s/n, E-38205, La Laguna, Tenerife, Spain\\
$^{6}$Dipartimento di Fisica, Universit\`a degli Studi di Torino, via Pietro Giuria 1, I-10125, Torino, Italy\\
$^{7}$Stellar Astrophysics Centre, Deparment of Physics and Astronomy, Aarhus University, Ny Munkegade 120, DK-8000 Aarhus C, Denmark\\
$^{8}$Department of Space, Earth and Environment, Chalmers University of Technology, Onsala Space Observatory, 439 92 Onsala, Sweden\\
$^{9}$IRFU, CEA, Universit\'e Paris-Saclay, F-91191 Gif-sur-Yvette, France\\
$^{10}$Universit\'e Paris Diderot, AIM, Sorbonne Paris Cit\'e, CEA, CNRS, F-91191 Gif-sur-Yvette, France\\
$^{11}$Th\"uringer Landessternwarte Tautenburg, Sternwarte 5, D-07778 Tautenberg, Germany\\
$^{12}$Department of Astronomy, Graduate School of Science, The University of Tokyo, Hongo 7-3-1, Bunkyo-ku, Tokyo, 113-0033, Japan\\
$^{13}$National Optical Astronomy Observatory, 950 North Cherry Avenue Tucson, AZ 85719, USA\\
$^{14}$Department of Earth and Planetary Sciences, Tokyo Institute of Technology, 2-12-1 Ookayama, Meguro-ku, Tokyo 152-8551, Japan\\
$^{15}$Max-Planck-Institut f\"ur Astronomie, K\"onigstuhl 17, D-69117 Heidelberg, Germany\\
$^{16}$School of Physics and Astronomy, University of Birmingham, Edgbaston, Birmingham, B15 2TT, UK\\
$^{17}$Department of Astronomy and McDonald Observatory, University of Texas at Austin, 2515 Speedway, Stop C1400, Austin, TX 78712, USA\\
$^{18}$Institute of Planetary Research, German Aerospace Center, Rutherfordstrasse 2, 12489 Berlin, Germany\\
$^{19}$Rheinisches Institut f\"ur Umweltforschung, Abteilung Planetenforschung an der Universit\"at zu K\"oln, Aachener Strasse 209,\\
50931 K\"oln, Germany\\
$^{20}$Astrobiology Center, NINS, 2-21-1 Osawa, Mitaka, Tokyo 181-8588, Japan\\
$^{21}$National Astronomical Observatory of Japan, NINS, 2-21-1 Osawa, Mitaka, Tokyo 181-8588, Japan\\
$^{22}$ZAH-Landessternwarte Heidelberg, K\"onigstuhl 12, 69117 Heidelberg, Germany\\
$^{23}$European Southern Observatory, Alonso de C\'ordova 3107, Vitacura, Casilla, 19001, Santiago de Chile, Chile\\
\vspace{-1.2cm}
}
\begin{document}
\label{firstpage}
\pagerange{\pageref{firstpage}--\pageref{lastpage}}
\maketitle

\begin{abstract}
We report the discovery and characterization of HD~89345b (K2-234b; EPIC~248777106b), a Saturn-sized planet orbiting a slightly evolved star.
HD~89345 is a bright star ($V = 9.3$ mag) observed by the K2 mission with one-minute time sampling. It exhibits solar-like oscillations. We conducted asteroseismology to determine the parameters of the star, finding the mass and radius to be $1.12^{+0.04}_{-0.01}~M_\odot$ and $1.657^{+0.020}_{-0.004}~R_\odot$, respectively. The star appears to have recently left the main sequence, based on the inferred age, $9.4^{+0.4}_{-1.3}~\mathrm{Gyr}$, and the non-detection of mixed modes.
The star hosts a ``warm Saturn'' ($P = 11.8$~days, $R_p = 6.86 \pm 0.14~R_\oplus$). Radial-velocity follow-up observations performed with the FIES, HARPS, and HARPS-N spectrographs show that the planet has a mass of $35.7 \pm 3.3~M_\oplus$. The data also show that the planet's orbit is eccentric ($e\approx 0.2$). An investigation of the rotational splitting of the oscillation frequencies of the star yields no conclusive evidence on the stellar inclination angle. We further obtained Rossiter-McLaughlin observations, which result in a broad posterior of the stellar obliquity.
The planet seems to conform to the same patterns that have been observed for other sub-Saturns regarding planet mass and multiplicity, orbital eccentricity, and stellar metallicity.
\end{abstract}
  


\section{Introduction}

When a planet transits, this opens up a potential window for dynamical studies (through e.g.\ the measurement of stellar obliquities) as well as atmospheric studies (through e.g.\ transmission spectroscopy), but unfortunately many host stars are too faint for these type of studies to be feasible.

We report the discovery and characterization of HD~89345b (K2-234b; EPIC~248777106b), a newly discovered transiting planet orbiting a bright star ($V = 9.3$) which was observed by the K2 mission \citep{howell2014}\footnote{During the reviewing stage of this manuscript, another manuscript independently reporting on this system was made publicly available \citep{yu2018}.}. This is a warm sub-Saturn planet. Such planets, with a size between Uranus and Neptune, do not exist in the solar system. They exhibit a wide variety of masses and their formation is not fully understood \citep{petigura2017}.

We confirm the existence of the planet and measure its mass with radial-velocity measurements, using the FIES \citep{telting2014}, HARPS \citep{mayor2003}, and HARPS-N \citep{cosentino2012} spectrographs. This work was done within the {\tt KESPRINT} collaboration \citep[see e.g.][]{sanchisojeda2015,vaneylen2016b,vaneylen2016a,gandolfi2017,fridlund2017,smith2018}. We determine accurate stellar parameters from asteroseismology, through the analysis of stellar oscillations that are seen in the {K2} light curve. 

In Section~\ref{sec:observations}, we describe the observations of this system, including the {K2} observations, high-resolution imaging, and spectroscopic observations. In Section~\ref{sec:star}, we describe the derivation of spectroscopic stellar parameters, and the asteroseismic analysis of the star. In Section~\ref{sec:planet}, we derive the properties of the planet and its orbit. We conclude with a discussion in Section~\ref{sec:discussion}.

\section{Observations}
\label{sec:observations}

\subsection{K2 photometry}
\label{sec:k2phot}

HD~89345 was observed by the {K2} mission \citep{howell2014} during Campaign 14 (UT May 31 to Aug 19, 2017). As a bright ($V=9.3$~mag) solar-type sub-giant star, HD~89345 was proposed as a short-cadence (with an integration time of 58.8 seconds) target to enable an asteroseismic analysis (Lund et al., guest observer program GO14010). We downloaded the target pixel files from the Mikulski Archive for Space Telescopes.\footnote{https://archive.stsci.edu/k2.} We first removed the systematic flux variation due to the rolling motion of the {\it Kepler} spacecraft. We 
adopted a similar procedure to that described by \cite{vanderburg2014}. In short, we put down a circular aperture around the brightest pixel in the target pixel files. We then fitted a two dimensional Gaussian function to the flux distribution within the aperture. The $x$ and $y$ positions of the Gaussian functions were used as tracers of the spacecraft's rolling motion. We fitted a piecewise linear function between the aperture-summed flux variation and the x and y positions. This function describes the systematic variation due to the rolling motion and was removed by division.

Prior to our transit detection, we removed any long-term astrophysical or instrumental flux variation by fitting a cubic spline to the light curve. We then searched the resultant light curves for periodic transit signals using the Box-Least-Square algorithm \citep{kovacs2002}. The signal of planet b was clearly detected with a signal-to-noise ratio (SNR) of 16. We searched for additional transiting planets after removing the transits of planet b. No additional signal was detected with SNR>4.5. 

\subsection{High-resolution photometry}
\label{sec:highresimage}

\begin{figure}
	\includegraphics[width=\columnwidth]{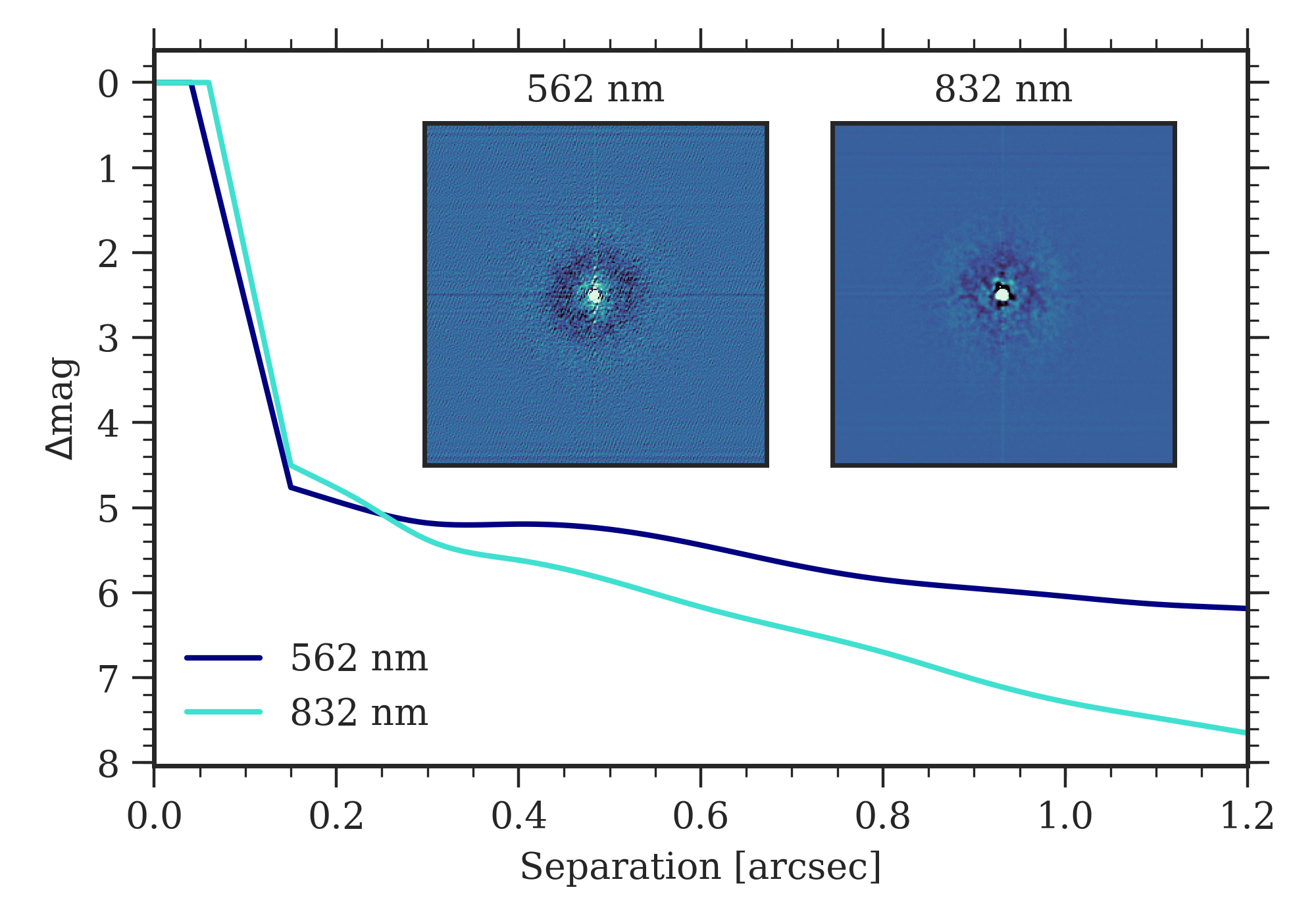}
	\caption{NESSI Speckle-interferometric observations of HD~89345 at 562nm and 832nm reveal no nearby stars. Contrast limits as a function of angular separation are shown (see Section~\ref{sec:highresimage} for details). The inset images have a scale of $4.6\arcsec\times4.6\arcsec$, and are oriented with northeast in the upper left.}
	\label{fig:highresimage}
\end{figure}

We conducted speckle-interferometry observations of the host star using the NASA Exoplanet Star and Speckle Imager \citep[NESSI,][Scott et al., in prep.]{scott2016} on the WIYN 3.5-m telescope. The observations were conducted at 562nm and 832nm simultaneously, using high-speed electron-multiplying CCDs with individual exposure times of 40~ms. The data were collected and reduced following the procedures described by \citet{howell2011}, resulting in reconstructed $4.6\arcsec\times4.6\arcsec$ images of the host star with a resolution close to the diffraction limit. 
We did not detect any secondary sources in the reconstructed images. We produced smooth contrast curves from the reconstructed images by fitting a cubic spline to the 5$\sigma$ sensitivity limits within a series of concentric annuli.
The achieved contrast of 5 mag at 0.2\arcsec~strongly constrains the possibility that a nearby faint star is the source of the observed transit signal. We show the reconstructed images and the resulting background source sensitivity limits in Figure~\ref{fig:highresimage}.

\subsection{Spectroscopic observations}
\label{sec:observations_spectro}

High resolution spectroscopic observations of HD~89345 were obtained between 23 December 2017 and 25 March 2018, using three different spectrographs.

Following the observing strategy described by \cite{gandolfi2013}, we gathered 16 high-resolution spectra (R\,=\,67000) of HD~89345 with the FIbre-fed \'Echelle Spectrograph \citep[FIES;][]{frandsen1999,telting2014} mounted at the 2.56m Nordic Optical Telescope (NOT) of Roque de los Muchachos Observatory (La Palma, Spain). The observations were carried out as part of our {K2} follow-up programs 2017B/059, 56-209, and 56-010. We reduced the data using standard Image Reduction and Analysis Facility (IRAF) and Interactive Data Language (IDL) routines and extracted the RVs via multi-order cross-correlations against the stellar spectrum with the highest SNR as a template.

We also acquired 38 spectra (program ID 0100.C-0808) with the HARPS spectrograph \citep[$R\,\approx\,115\,000$;][]{mayor2003} mounted at the ESO-3.6m telescope of La Silla observatory (Chile), as well as 12 spectra (program IDs 2017B/059, A36TAC\_12, and CAT17B\_99) with the HARPS-N spectrograph \citep[$R\,\approx\,115\,000$;][]{cosentino2012} mounted at the 3.6m Telescopio Nazionale Galileo (TNG) of Roque de los Muchachos Observatory. To account for possible RV drifts of the instruments we used the simultaneous Fabry Perot calibrator. In the attempt to measure the sky-project spin-orbit angle, $\lambda$, 21 HARPS spectra were gathered during the transit occurring on the night 23/24 February 2018. We reduced the data using the dedicated offline HARPS and HARPS-N pipelines and extracted the RVs via cross-correlation with a numerical mask for a G2 type star.

In order to detect the transiting planet in the Doppler observations and exclude false positive scenarios (e.g., a background binary) we performed a frequency analysis of the RVs and their activity indicators (BIS and FWHM). On epochs 2458129 and 2458140 we collected FIES and HARPS-N spectra of HD\,89345 within about 1 hour. Similarly, on epochs 2458143 and 2458144 we obtained FIES and HARPS data within about 2 hours. We used these measurements to estimate the offsets of the RV, FWHM, BIS between the instruments and calculate the periodograms of the combined data. These offsets have only been used to perform the frequency analysis. For the procedure of the joint RV modeling, we refer the reader to Sect.\,4.

\begin{figure}
	\includegraphics[width=\columnwidth]{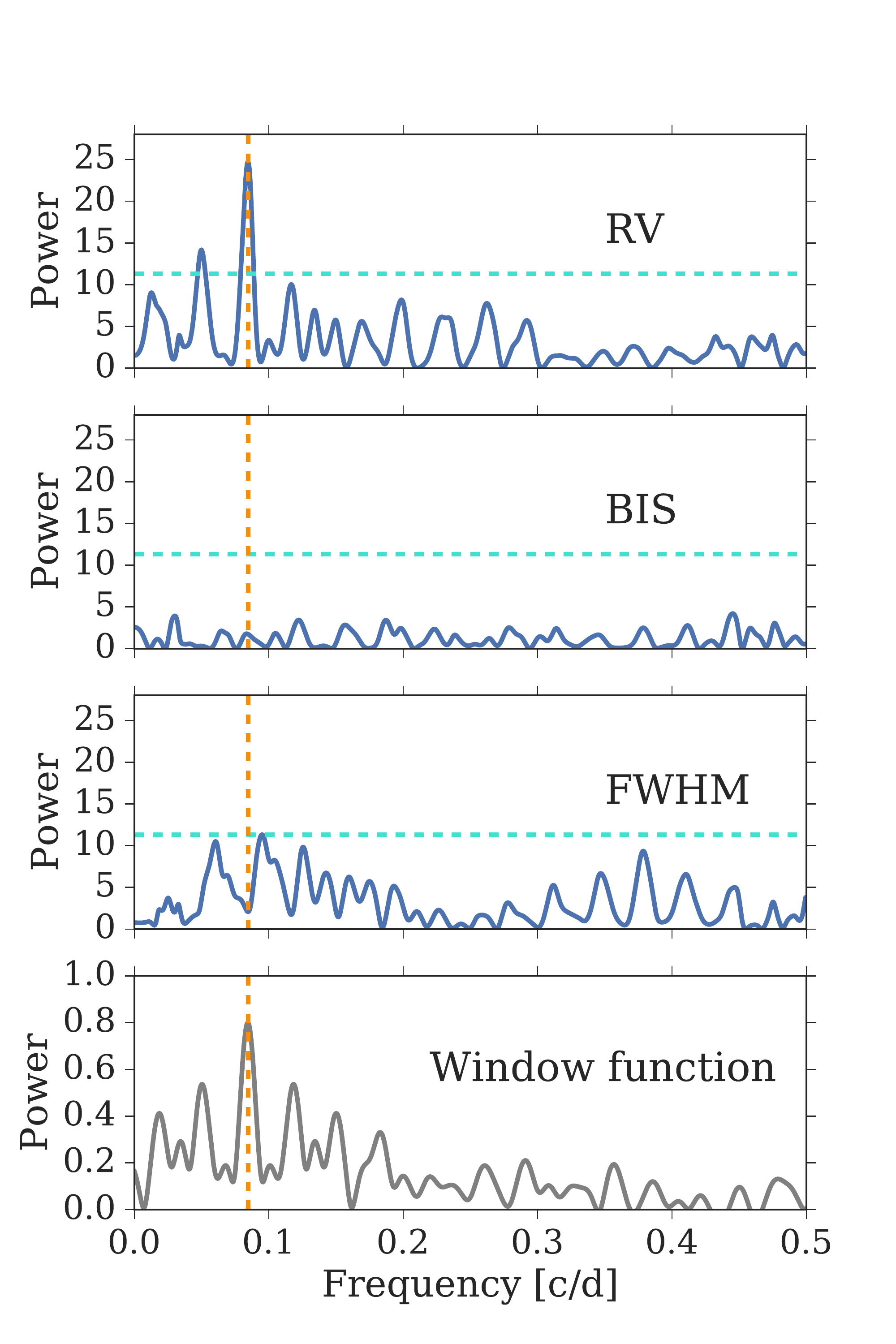}
	\caption{Generalized Lomb Scargle periodogram of the RV, BIS, and FWHM measurements for the combined FIES, HARPS, and HARPS-N observations, and the window function centered at the orbital frequency of the transiting planet. The RV peak at the orbital period observed from transit observations (vertical orange dotted line) does not have a corresponding BLS or FWHM peak, suggesting that it is induced by the planet. The light blue dotted horizontal line indicates a 0.01\% false alarm probability.}
	\label{fig:gls}
\end{figure}

The first three panels of Figure~\ref{fig:gls} display the generalized Lomb-Scargle periodograms \citep{zechmeister2009} of the combined RV, BIS, and FWHM measurements. The dashed vertical line marks the orbital frequency of the transiting planet, whereas the horizontal lines represent the 0.01\,\% false-alarm probability (FAP). We determined the FAP following the Monte Carlo bootstrap method described in \citet{kuerster1997}. In the last panel, we show the GLS of the window function shifted to the right by 0.085~c/d (i.e., the orbital frequency of the transiting planet), and mirrored to the left of this frequency, to facilitate visual identification of possible aliases.

The periodogram of the RV measurements has a strong peak at the orbital frequency of the transiting planet with a FAP\,$\ll$\,0.01\,\%, implying that we would infer the presence of the transiting planet even in the absence of K2 photometry. This peak has no counterparts in the periodograms of the BIS and FWHM, suggesting that the observed RV variation is induced by the transiting planet. We note the periodogram of the RV displays peaks separated by about 0.034\,c/d, which corresponds to about 30 days. Those peaks are aliases of the planet's frequency and are due to the fact that our observations have been performed around new moon to avoid contamination from the scattered Sun light.

All RV data points and their observation times are listed in Table~\ref{tab:rvdata}, along with the BIS, FWHM, exposure times and SNR per pixel at 5500\,\AA. For the HARPS and HARPS-N data, we also report the activity index $\log R'_\mathrm{HK}$ of the Ca\,{\sc ii} H\,\&\,K lines.

\section{Stellar parameters}
\label{sec:star}	

We determined the stellar parameters based on spectroscopy, parallax and magnitude measurements, and asteroseismology. Below we describe each of these methods. We also investigated the inclination angle of the star based on rotational splittings of the oscillation modes.

\subsection{Spectroscopic analysis}
\label{sec:spectroscopic}

\begin{table}
\begin{center}
\caption{Spectroscopic parameters (see Section~\ref{sec:spectroscopic}). \label{tab:spectroscopic}}
    \begin{tabular}{l  r@{$\pm$}l }
      \hline \hline
      Parameter & \multicolumn{2}{c}{Value}\\
      \hline
      Effective Temperature, $T_{\rm_{eff}}$ (K) 			& $5420$ & $110$ 	\\
      Surface gravity from \mgi, $\log g$ (cgs)       			& $3.85$ & $0.20$ 	\\
      Surface gravity from \cai, $\log g$ (cgs)	      			& $3.85$ & $0.13$ 	\\
      Metallicity, [Fe/H]                        			& $0.45$ & $0.05$ 	\\
      Projected rotation speed, $v \sin i$ (km\,s$^{-1}$) 	& $2.60$ & $0.50$  \\
      Microturbulence (km\,s$^{-1}$)             			& $0.80$ & $0.10$ 	\\
      Macroturbulence (km\,s$^{-1}$) 					& $3.51$ & $0.50$  \\
\end{tabular}
\end{center}
\end{table}

In order to derive the stellar parameters, we combined all the HARPS spectra (see Section~\ref{sec:observations_spectro}) to form a co-added spectrum with a SNR of about 500 per pixel at 5500~\AA. This was analysed using the spectral analysis package Spectroscopy Made Easy \citep[SME,][]{valenti1996,valenti2005,piskunov2017}. 
SME calculates synthetic spectra in local thermodynamic equilibrium (LTE) for a set of given stellar parameters and fits them to observed high-resolution spectra using a $\chi^2$  minimisation procedure. We used SME version 5.2.2 and a grid of the ATLAS12 model atmospheres \citep{kurucz2013}, which is a set of one-dimensional (1D) models applicable to solar-like stars.
 
We then fitted the observed spectrum to this grid of theoretical ATLAS12 model atmospheres, selecting parts of the observed spectrum that contain spectral features that are sensitive to the required parameters. We used the empirical calibration equations for solar-like stars from \cite{bruntt2010}, in order to determine the micro-turbulent and macro-turbulent velocities, respectively. We then followed the procedure in \cite{fridlund2017}. In short, we used the wings of the Hydrogen Balmer lines to determine the effective temperature, \teff\, \citep{fuhrmann1993,fuhrmann1994}. The line cores were excluded in this fitting procedure due to their origin in layers above the photosphere. 

The stellar surface gravity, \logg\, was estimated from the wings of the \cai\ 6102, 6122, 6162 triplet, and the \cai\ 6439 \AA\ line. 
We separately determined \logg\ from the \mgi\ 5167, 5172, 5183 triplet and found a result consistent within $1\sigma$. We conservatively adopted the value from \mgi, which has the highest uncertainty.

The projected stellar rotational velocity, \vsini, and the metal abundances, were measured by fitting the profile of several tens of clean and unblended metal lines. The final model was checked with the Na doublet (5889 and 5896 \AA).
The velocity profile of the absorption lines have a degeneracy caused by the combination of the macro turbulence ($V_\mathrm{mac}$) and the rotational velocity component, \vsini. Although there are theoretical models for $V_\mathrm{mac}$, empirical calibrations have been made by \cite{bruntt2010} and \cite{doyle2014}. Both use a combination of spectroscopic and asteroseismic analysis in order to correlate macro-turbulence and rotation for a sample of about 50 stars. While the number of stars in each sample (about 25) is relatively small, together they demonstrate clearly empirical trends which can be used to assign a value to $V_\mathrm{mac}$ after $T_\mathrm{eff}$ has been determined. In the case of this star, there is a small difference between both calibrations. The relation by \cite{bruntt2010} indicates $V_\mathrm{mac} = 1.7 \pm 0.4$~km~s$^{-1}$, while using the relation of \cite{doyle2014} results in $V_\mathrm{mac} = 3.51 \pm 0.5$~km~s$^{-1}$. This leads to \vsini\ of $3.45 \pm 0.50$ and $2.60 \pm 0.50$~km~s$^{-1}$, respectively, for the two values of $V_\mathrm{mac}$. Here, we adopt the calibration by \cite{doyle2014} for two reasons. Firstly, the treatment of the asteroseismic data is more thorough in this work, since it had access to high-quality data from the \textit{Kepler} space mission, which allowed them to dig deeper into the rotational aspects of the target stars. Secondly, the values for the empirical sample of \cite{bruntt2010} tend to be lower than values by \cite{doyle2014}, but also lower than data by \cite{gray1984} and \cite{valenti2005}. The latter used the SME modeling tool, that we have also used to interpret our spectroscopic data here. Finally, we note that the lower \vsini\ value is also more consistent with limits derived from in-transit spectroscopic observations (see Section~\ref{sec:rm}).

All spectroscopic parameters are listed in Table~\ref{tab:spectroscopic}. 

\subsection{Parallax measurements}
\label{sec:parallax}

\begin{table}
\caption{We list the GAIA parallax measurement, as well as magnitude measurements in different colors, and the stellar parameters we derived from these observations (see Section~\ref{sec:parallax}).  \label{tab:parallax}}
\begin{center}
\begin{tabular}{ cccccccccc } 
\hline \hline
Parameter & Value & Source\\
\hline
Paral.\ [mas]	& $7.527 \pm 0.046$ & \cite{gaia2018} \\
B Mag.\ & $10.15 \pm 0.04$ &	\cite{hog2000} \\ 
V Mag.\ & $9.38 \pm 0.03$ &	 \cite{hog2000} \\
G Mag.\ & $9.159 \pm 0.001$ &	\cite{gaia2016} \\ 
J Mag.\ & $8.091 \pm 0.020$ &	\cite{cutri2003} \\ 
H Mag.\ & $7.766 \pm 0.040$ &	\cite{cutri2003} \\
K Mag.\ & $7.721 \pm 0.018$ &	\cite{cutri2003} \\ 
\hline
$R$ $[R_{\odot}]$ 	& $1.78^{+0.06}_{-0.06}$ & This work \\
$M$ $[M_{\odot}]$ 	& $1.10^{+0.06}_{-0.14}$ & This work \\
$L$ $[L_\odot]$		& $2.71^{+0.12}_{-0.12}$ & This work \\
\end{tabular}
\end{center}
\end{table}

We use the parallax and the observed apparent magnitudes to obtain an independent estimate of the stellar parameters. This was done using {\tt BASTA} \citep{silvaaguirre2015} with a grid of BaSTI isochrones \citep{pietrinferni2004}. 
The BaSTI isochrones contain synthetic colors and absolute magnitudes in a range of photometric broadband filters. Using the Gaia parallax \citep[see Table~\ref{tab:parallax}][]{gaia2016a,gaia2018}, we convert apparent magnitudes to absolute magnitudes. Following \cite{luri2018}, we add 0.1 mas in quadrature to the uncertainty of the parallax, to account for systematic uncertainty. We estimate the reddening $E(B-V)$ along the line-of-sight using the \cite{green2015} dust map and transform $E(B-V)$ to extinction $A_{\lambda}$ in different filters following \cite{casagrande2014}. 
The extinction-corrected absolute magnitudes are fitted to the grid of isochrones following the Bayesian grid-modeling approach employed by {\tt BASTA}. We fitted the Johnson $V$ and $B$ magnitudes as well as 2MASS $J$, $H$ and $K$ magnitudes and derive the stellar luminosity, mass and radius. All parameters are listed in Table~\ref{tab:parallax}. 

\subsection{Asteroseismic analysis}
\label{sec:analysis_asteroseismic}

We subsequently determined stellar parameters using asteroseismology. The A2Z pipeline \citep{mathur2010} was used on the reduced {K2} photometry (see Section~\ref{sec:k2phot}), after excising the data obtained during transits. The pipeline determines the global seismic parameters $\Delta \nu$, the mean large frequency spacing, and $\nu_{\rm max}$, the frequency of maximum power. The first parameter is given by the distance in frequency between two modes of the same angular degree and of consecutive orders, a quantity which is proportional to the square root of the mean density of the star \citep{kjeldsen1995}. The frequency of maximum power is related to the cut-off frequency, which is directly proportional to the surface gravity of the star \citep{brown2011}. This resulted in a first estimate of the global seismic parameters for this star: $\Delta \nu$=67.00\,$\pm$\,1.87\,$\mu$Hz and $\nu_{\rm max}$=1300\,$\pm$\,58\,$\mu$Hz.

\begin{figure*}	
	\includegraphics[width=2\columnwidth]{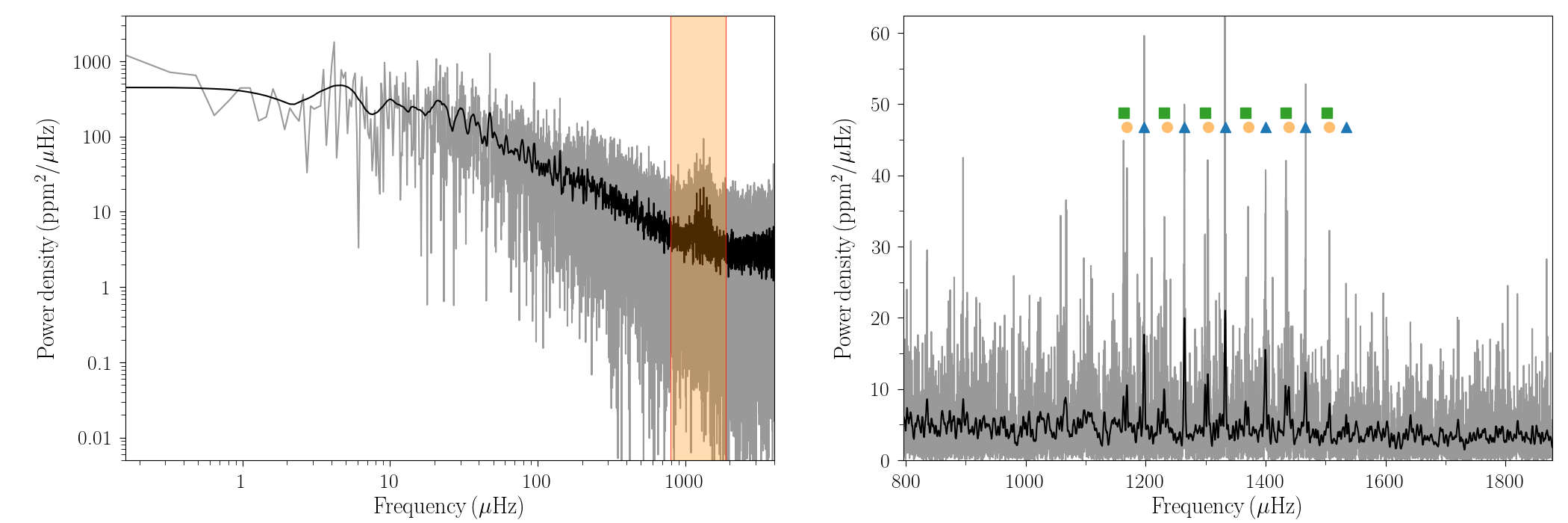}
	\caption{Power spectrum density of HD~89345, showing a power excess due to stellar oscillations, based on the {K2} photometry. The right panel is a zoom-in on the region with solar-like oscillations. The power spectrum is shown in grey and a smoothed version is shown in black. The color symbols indicate the derived frequencies as listed in Table~\ref{tab:frequencies}.}
	\label{fig:powerspectrum}
\end{figure*}

We determined the set of individual $p$-mode frequencies using two methods. The first method involves maximum {\it a priori} (MAP) fitting.
To reduce the number of free parameters, all the modes with $l=0$, $l=1$, and $l=2$ were fitted together \citep{rocacortes1999}, assuming one single Lorentzian profile per mode (without accounting for any rotation), a constant line width and amplitude per order, and constant visibilities between the modes ($1$, $1.5$, and $0.5$, respectively, for $l=0$, $1$ and $2$). To validate this last assumption we also fitted the data leaving the visibilities as free parameters, and found that the result of this fit agrees with the constant visibilities to within the uncertainties, as do the fitted mode frequencies. The {K2} photometry used in this analysis were treated with the KADASC correction pipeline \citep{garcia2011}. The transits were removed and the data were interpolated using inpainting methods \citep{garcia2014,pires2015}.

The second frequency extraction method uses the Bayesian methodology outlined by \citet{lund2017}, which was applied to data prepared using the K2P$^2$ pipeline to extract and correct the {K2} photometry in a way that is optimal for determining oscillation frequencies \citep{lund2014,lund2016}.

The frequencies of these two methods agree to within the estimated $1\sigma$ uncertainties for all frequencies.  The MAP fitting identified additional low-amplitude frequency detections. We adopt the frequencies provided by the Bayesian method for the modeling, because this methodology provide access to the posterior probabilities of each fitted parameter. A power spectrum of the {K2} photometry is shown in Figure~\ref{fig:powerspectrum}, together with the detected Bayesian frequencies. We list all frequencies in Table~\ref{tab:frequencies}. 

We subsequently modeled the oscillation frequencies following two different approaches. The first stellar modeling method makes use of the MESA evolution code \citep{paxton2011}. 
The OPAL opacities \citep{iglesias1996}, the GS98 metallicity mixture \citep{grevesse1998}, and the exponential prescription of \cite{herwig2000} for the overshooting were used, and otherwise the standard input physics from MESA was applied. 
The frequencies of the acoustic modes were calculated with the ADIPLS code \citep{christensendalsgaard2008} in the adiabatic approximation. 
A $\chi^2$ minimization including $p$-mode frequencies and spectroscopic data was applied to a grid of models. The general procedure is described in \citet{perez2016}. 
However, since HD~89345 is a subgiant star with eigenfrequencies approximately in the asymptotic $p$-mode regime, all the
modes given in Table~\ref{tab:frequencies} were fitted simultaneously with weights based on their observational errors and the same surface correction was applied to all the modes, i.e.\ a second order polynomial fit to the relative differences $I_{nl}\delta \omega_{nl}/\omega_{nl}$ , where $I_{nl}$ is the dimensionless energy \citep[see][for more details]{perez2016}.
The input spectroscopic parameters considered were the effective temperature, surface gravity, and metallicity (see Table~\ref{tab:spectroscopic}). 
The grid is composed of evolution sequences with stellar masses ($M_\star$) from $0.95~M_{\odot}$ to $1.25~M_{\odot}$ with a step of $\Delta M = 0.01~M_{\odot}$, initial metalicities ($Z_{\mathrm{ini}}$) from $0.002$ to $0.04$ with a step of $\Delta Z=1/300$, mixing length parameters ($\alpha$) from $1.5$ to $2.2$ and step $\Delta \alpha=0.1$ and overshooting parameter $f_{\rm ov}$ from $0$ to $0.04$ and step of $0.01$. 
The helium abundance was constrained by adopting a Galactic chemical evolution model with $\Delta Z/\Delta Y=1.4\,$. 

To estimate the uncertainty in the output parameters we assumed normally distributed uncertainties for the observed frequencies, for the mean value of $I_n\delta \omega/\omega$ for radial oscillations and for the spectroscopic parameters. We then search for the model with the minimum $\chi^2$ in every realization, and report mean and 1$\sigma$ uncertainty values in Table~\ref{tab:asteroseismology}.

\begin{table}
\caption{Stellar parameters derived from asteroseismic modeling using two different approaches (see Section~\ref{sec:analysis_asteroseismic}).}
\label{tab:asteroseismology}
\centering
\begin{tabular}{c c c c c c c c}
\hline \hline
Parameter 		& MESA 				& BASTA \\
\hline
 $R$ $[R_{\odot}]$ 	& $1.657  \pm 0.017$ 		& $1.657^{+0.02}_{-0.004}$ \\ [3pt]
 $M$ $[M_{\odot}]$ 	& $1.11\pm 0.04$ 		& $1.120^{+0.04}_{-0.01}$ \\ [3pt]
 $\rho$ [g/cm$^3$] 	& $0.3413\pm 0.0010$ 		& $0.343 \pm 0.002$ \\
 $T_\mathrm{eff}$ [K]	& $5480 \pm 100$		& $5499 \pm 73$ \\
$L [L_\odot]$		& $2.21 \pm 0.22$		& $2.27^{+0.21}_{-0.14}$ \\ [3pt]
 Age [Gyr] 		& $8.3\pm 1.2$  		& $9.4^{+0.4}_{-1.3}$	\\ [3pt]
 $\log g$ [dex]		& $4.045\pm 0.007$		& $4.044^{+0.006}_{-0.004}$ \\ [3pt]
 $\alpha$ 		& $1.53 \pm  0.06$  		& 1.7917 (fixed) \\
 $f_{\rm{ov}}$ 		&  $0.004 \pm 0.007$ 		& 0 (fixed) \\
\end{tabular}
\end{table} 

\begin{figure}
	\includegraphics[width=\columnwidth]{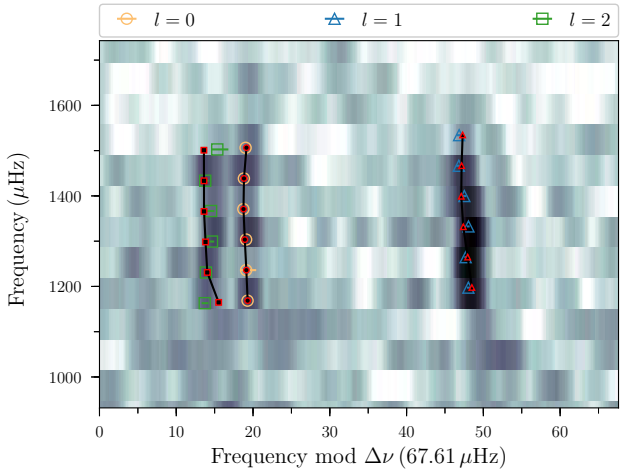}
	\caption{\'Echelle diagram of HD~89345, showing the observed power as a function of frequency and frequency modulus the large frequency separation. The determined frequencies are shown in different colors and listed in Table~\protect\ref{tab:frequencies}, and the best model frequencies from {\tt BASTA} are overplotted with red symbols connected by black lines.}
	\label{fig:echelle}
\end{figure}

In the second approach, we made use of the BAyesian STellar Algorithm \texttt{BASTA} \citep{silvaaguirre2015}. \texttt{BASTA} uses a Bayesian grid-modelling approach and fits spectroscopic and asteroseismic observables to a large grid of stellar models. We used the grid of stellar models constructed for the \textit{Kepler} LEGACY sample \citep{lund2017,silvaaguirre2017}. The grid is built using GARSTEC evolutionary models \citep{weiss2008} with oscillation frequencies computed using ADIPLS \citep{christensendalsgaard2008}. We used the OPAL05 equation-of-state \citep{rogers2002}, the GS98 solar mixture \citep{grevesse1998} and OPAL96 \citep{iglesias1996} and \cite{ferguson2005} opacities. The inclusion of microscopic diffusion or overshooting does not significantly affect the derived parameters. We fitted the spectroscopically derived $T_{\mathrm{eff}}$, $\log g$ and [Fe/H] and the frequency ratios $r01$, $r10$ and $r02$. We fit frequency ratios \citep[as defined by][]{roxburgh2013} since these are less affected by the asteroseismic surface effect than individual oscillation frequencies, which need corrections to match theoretical frequencies. We report $16\%$, $50\%$, and $84\%$ percentile values from $\texttt{BASTA}$'s probability distributions. 
All stellar parameters are listed in Table~\ref{tab:asteroseismology}.

As can be seen in this table, there is good agreement between the stellar parameters derived from the two frequency modeling approaches. Both sets of parameters also agree well with the spectroscopic parameters (see Table~\ref{tab:spectroscopic}), some of which were used as a prior in the asteroseismic modeling, and the parameters derived from parallax and color information (see Table~\ref{tab:parallax}). The asteroseismic radius and mass have a precision of $1.2\%$ and $3.6\%$, respectively, which are significantly more precise than the parallax measurements (with a precision of $3.3\%$ and $13\%$, respectively) and than what can typically be achieved with spectroscopy.

To calculate planetary parameters, we adopt the {\tt BASTA} stellar parameters, which have been previously used and tested for exoplanet host stars \citep[e.g.][]{davies2015,silvaaguirre2015,lund2017,silvaaguirre2017}. We show the frequencies of the best {\tt BASTA} model in Figure~\ref{fig:echelle}, together with the observed frequencies.

\subsection{Stellar rotation and inclination}
\label{sec:splitting}

\subsubsection{Asteroseismic analysis}
\label{sec:asteroseismic_splitting}

\begin{figure}
    \includegraphics[width=\columnwidth]{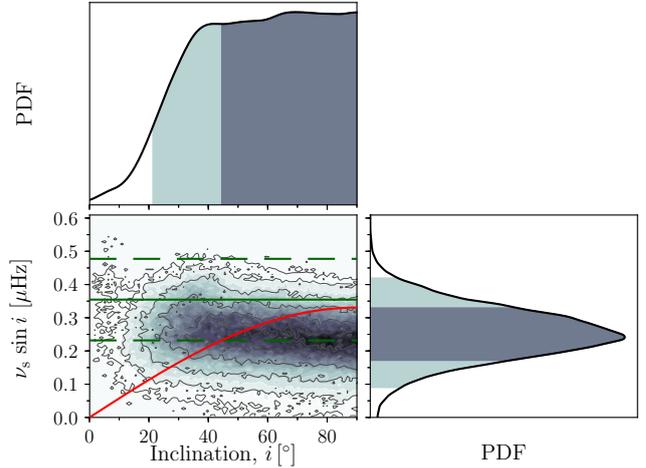}
    \caption{
    Posterior distribution of the stellar inclination versus the projected rotational splitting of the oscillation frequencies. The splitting of the frequencies is related to $\nu\sin i$ and subject to a Gaussian prior based on the measured projected rotational velocity v$\sin i$ and stellar radius (green lines in plot), while the relative amplitudes of the split frequencies provide information about the stellar inclination (see Section~\ref{sec:splitting}). The red line indicates the projected splitting corresponding to a rotation period of $35$ days, as found from analysis of the light curve. In dark blue and light blue, the 68\% and 95\% highest probability density
    intervals are indicated, respectively.
    }
    \label{fig:splitting}
\end{figure}

As part of the Bayesian frequency determination \citep{lund2017} described above, we also modeled the splitting of oscillation frequencies under the influence of rotation \mbox{\citep{gizon2003,ballot2006}}. In some cases, the rotational splitting can provide both the stellar rotation rate and its inclination, leading to a constraint on the obliquity of stars that host transiting planets \citep[see e.g.][]{chaplin2013,vaneylen2014,lund2014,campante2016}.
\clearpage 

Specifically, we modeled the projected splitting ($\nu_\mathrm{s} \sin i$, with $\nu_\mathrm{s}$ the observed frequency splitting and $i$ the stellar inclination) using prior constraints based on the previously determined stellar radius and the spectroscopic v$\sin i$ value.  We also tried modeling the splitting without these prior constraints. In both cases the overall result for the inclination is the same, but the best constraint is achieved when using a prior on v$\sin i$ and stellar radius, which corresponds to a prior on the projected rotational splitting of $\rm 0.35\pm 0.13\,  \mu$~Hz [$\nu\sin i$ = (v$\sin i)/2\pi R$], and further placing a uniform prior on the cosine of the stellar inclination. As shown in Figure~\ref{fig:splitting} the inclination is consistent with an aligned orbit, i.e., $i = 90^\circ$, and can at the $1\sigma$ limit only be constrained to a lower value of $i \geq 44^\circ$.

The uncertainty is caused by the relatively short duration of {K2} photometry. Seismic analysis done with CoRoT \citep{baglin2006} have placed a limit in the minimum length necessary to have reliable measurements of the inclination angle in G- and K-type stars at about 100 continuous days \mbox{\citep[e.g.][]{gizon2013,mathur2013}}.
Using \textit{Kepler}, precise inclination measurements have been measured using several years of observations for many stars, including some stars hosting transiting planets {\citep[e.g.][]{chaplin2013,huber2013obliquity,vaneylen2014,campante2016}}.
%
We also inspected the {K2} light curve for signatures of surface rotation following the methods described in \cite{garcia2014b}. A signal was detected at around 35 days, but due to the short timespan of the observations ($\approx 80~$days) it is difficult to confirm that this periodicity is indeed the rotation period of the star. We note, however, that a rotation period of 35 days is consistent with the estimated v$\sin i$ from spectroscopy and with the estimated projected splitting of ${\sim}0.25\pm0.1\,  \mu~$Hz at a stellar inclination above the  $1\sigma$ lower limit (see Figure~\ref{fig:splitting}).

\subsubsection{Rossiter-McLaughlin observations}
\label{sec:rm}

\begin{figure}
	\includegraphics[width=\columnwidth]{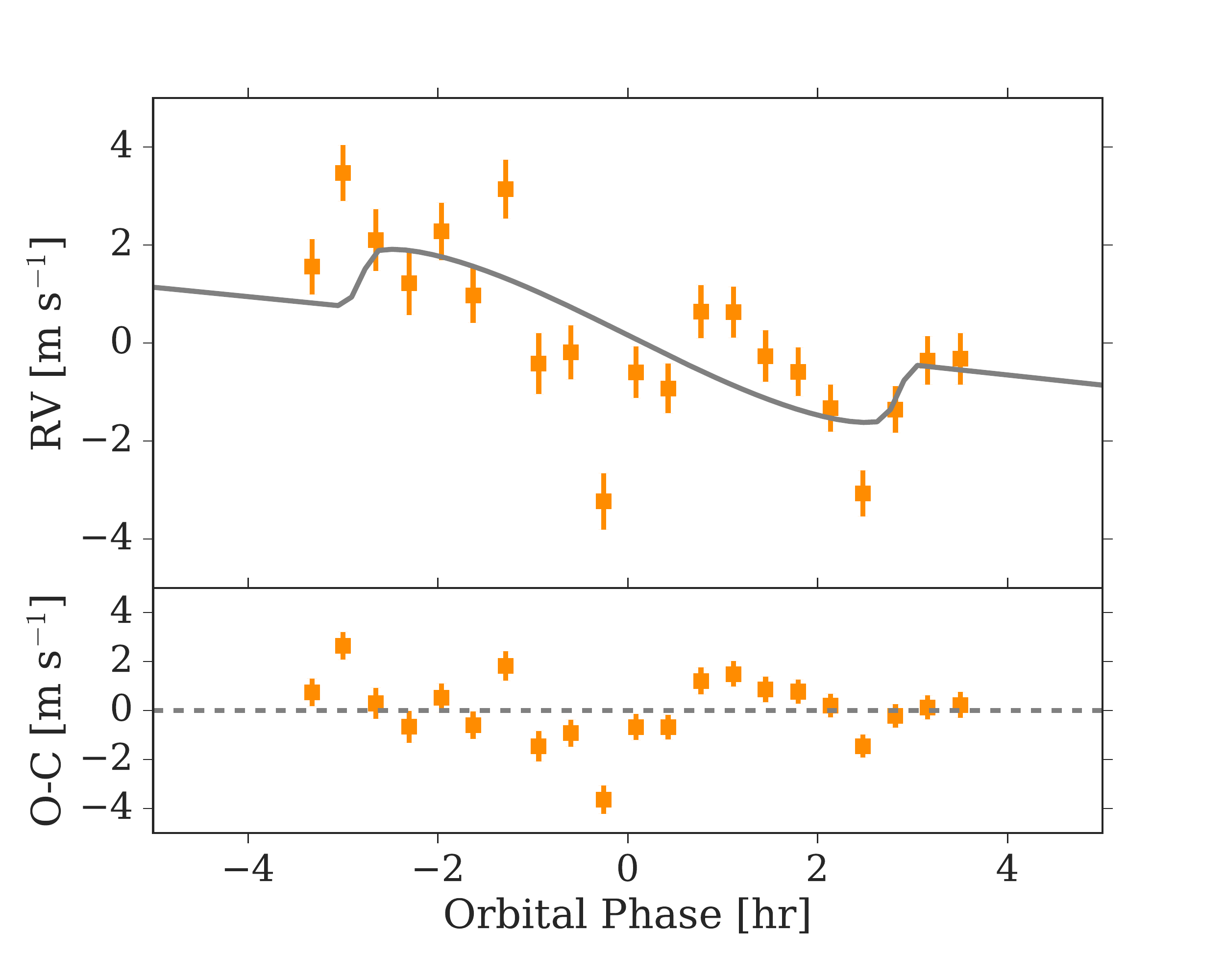}
    \caption{In-transit RV observations measured on the night of 23/24 February 2018 using HARPS. The top panel shows the observations and the best-fitting model of the Rossiter-McLaughlin effect are plotted, as described in Section~\ref{sec:rm}, and the bottom panel shows the residuals.}
    \label{fig:RM}
\end{figure}

Using in-transit spectroscopic observations (see Section~\ref{sec:observations_spectro}), we modeled the Rossiter-McLaughlin \citep[RM][]{rossiter1924,mclaughlin1924} effect following the approach of \cite{albrecht2012} and using the code of \cite{hirano2011} assuming solid body rotation of the stellar photosphere. 

Besides $\lambda$, the following model parameters were fitted: $v \sin i$, the limb darkening parameters, $u_1$ and $u_2$, the planet-to-star radius ratio, $R_p/R_{\star}$, the time of mid-transit, $t_c$, the scaled orbital distance, $a/R_{\star}$, the RV semi-amplitude of the star, $K_{\star}$, the systemic velocity of HARPS, $\gamma_{\text{HARPS}}$, as well as the orbital inclination $i$, and parameters representing the microturbulence $\beta$ and macroturbulence $\zeta$. The results from the joint planet modeling (see Section~\ref{sec:planet} and Table~\ref{tab:parameters}) were used as priors on all parameters except for $\lambda$, $v \sin i$ and $\gamma_{\text{HARPS}}$. The analysis was done for fixed values of $P$, $e$ and $\omega$, since these have minimal influence of the shape of the RM signal. We solved for the best-fit solution for the parameters and their posterior distribution using an MCMC analysis with {\tt emcee} \citep{foremanmackey2013}. We initialized 120 walkers in the vicinity of the best-fit solution. We ran the walkers for 1500 steps and discarded the first 800 steps as the burn-in phase.

As can be seen in Figure~\ref{fig:RM}, the data shows no clear RM signal. We find \vsini = $1.4^{+1.1}_{-0.8}$~km~s$^{-1}$, which is consistent with the value derived from spectroscopic analysis (see Section~\ref{sec:spectroscopic}). We further find $\lambda = 2^{+54}_{-30}$~degrees, consistent with alignment, but also with a broad range of obliquities, making it difficult to make conclusive statements about the stellar obliquity. 

We caution the reader against over interpreting this result. As discussed by \cite{albrecht2011} and \cite{triaud2017}, low SNR detections of the RM effect can lead to spuriously significant results for the projected obliquity. 
The apparently statistically significant result for lambda is based on RV data which appears to have not a significantly higher deviation from the orbital solution -- without the modelling of the RM effect -- than the out of transit data (see Figure~\ref{fig:RM}). If a clear detection of the RM effect was made, this would be the case.
However, a transit has occurred so two additional free parameters (\vsini\ and $\lambda$) are fitted for, but the RM measurement could
be the result of a particular realisation of measurement noise. Modeling the data with a systemic velocity ($\gamma$) and the orbital velocity ($K_\star$) does in effect apply a high pass filter. The functional forms of the RM effect for 90~deg and -90~deg orbits have a lower frequency than prograde and retrograde orbits, potentially leading to a spurious result in $\lambda$\footnote{We note that if the system would have a low impact parameter (which is not the case here) then the RM signal could be suppressed by having polar orbits ($|\lambda| \approx~90$~deg) and potential biases for a low-SNR RM measurement would differ.} Furthermore, the RM amplitude for projected obliquities near 90 deg and -90 deg is larger than for 0~deg and 180~deg orbits. 
This is because the maximum RV amplitude of the stellar photosphere which is covered by the transiting planet, and the lowest level of stellar limb darkening, occur during the same phase for the latter case, but not for the former case \citep[see][for details]{albrecht2013}.
Taking all this together, we conclude that additional measurements are needed to securely measure the projected obliquity in this system.

\section{Orbital and planetary parameters}
\label{sec:planet}

\subsection{Transit Model}
To model the transit light curve, we used the {\tt Python} package {\tt Batman} \citep{kreidberg2015}.  We isolated each transit with a 10-hour window around the time of mid-transit.  The transit model contains the following parameters: the orbital period $P_{\text{orb}}$, the mid-transit time $t_{\text{c}}$, the planet-to-star radius ratio $R_{\text{p}}/R_\star$; the scaled orbital distance $a/R_\star$; and the impact parameter $b\equiv a\cos i/R_\star$, and we adopted the quadratic limb-darkening profile, with parameters $u_1$ and $u_2$.

\begin{figure}
	\includegraphics[width=\columnwidth]{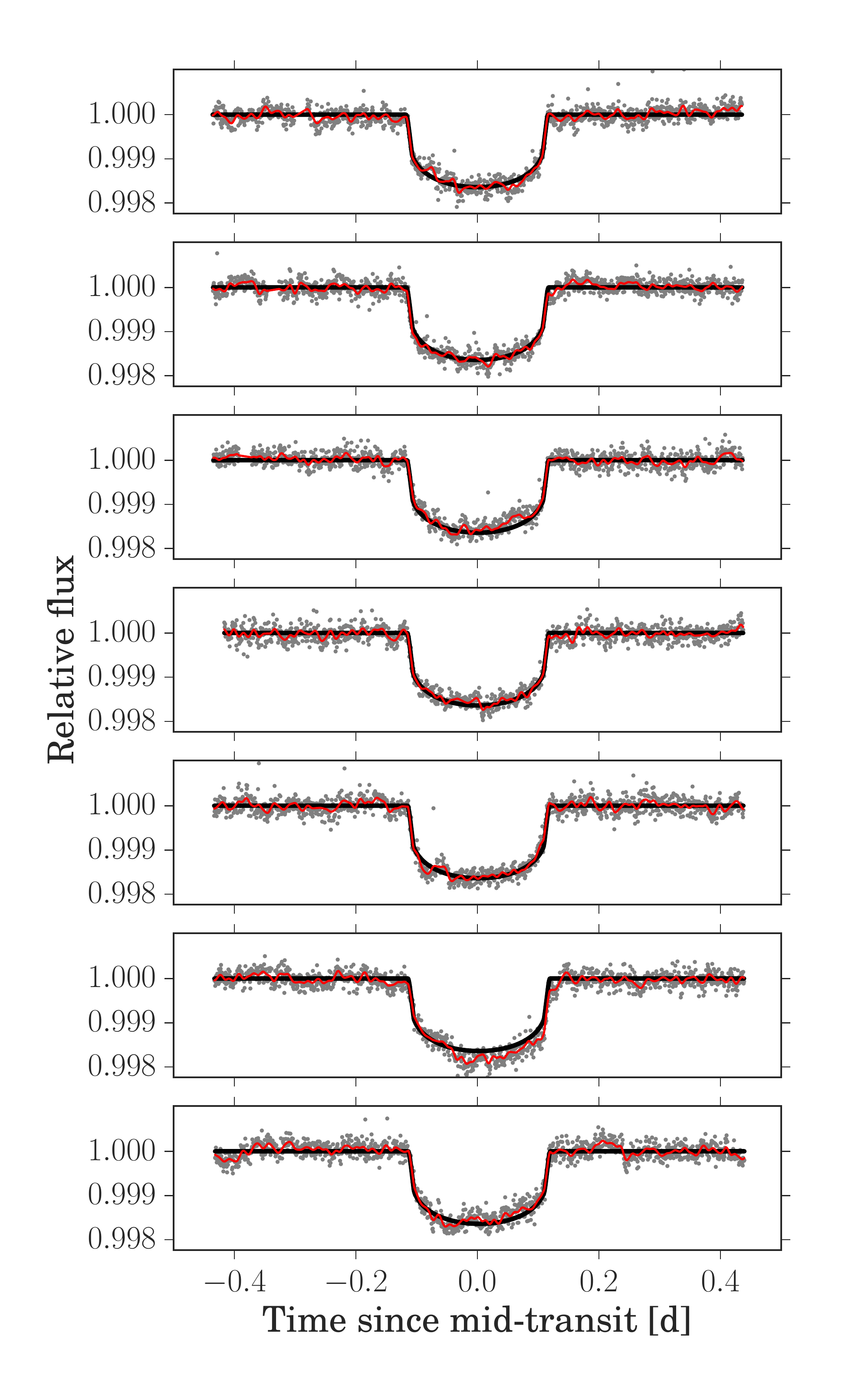}
	\caption{The transits observed with {K2} are shown in grey. Overplotted is the best transit model (black) and the best transit model including Gaussian processes (red), for the eccentric fitting case (see Section~\ref{sec:eccentricity}).}
	\label{fig:transit}
\end{figure}

\begin{figure}
	\includegraphics[width=\columnwidth]{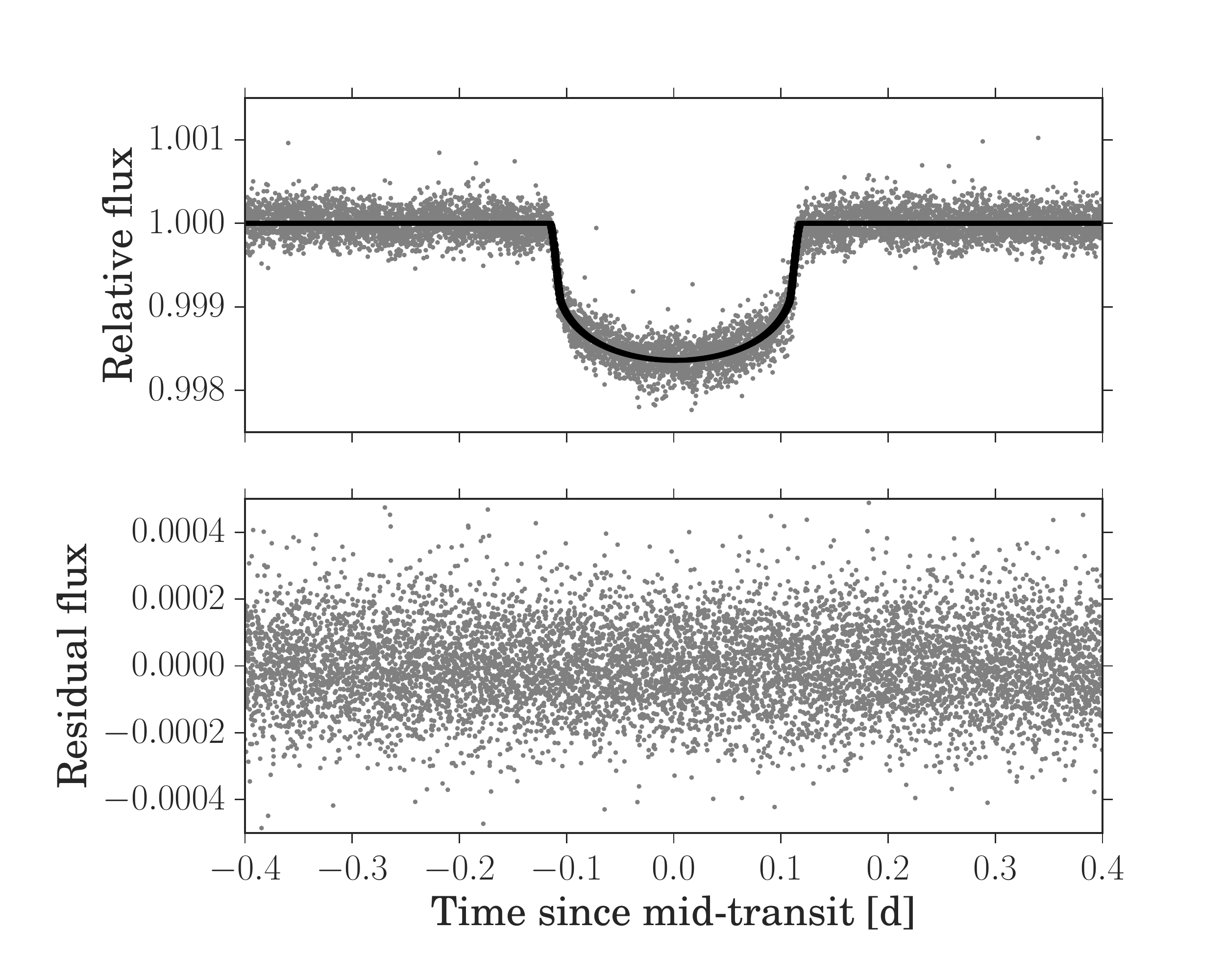}
	\caption{Combined {K2} transits (grey) together with the best-fitting model (black), not taking into account the Gaussian processes. The bottom panel shows the residuals.}
	\label{fig:combotransit}
\end{figure}

\subsection{Gaussian Process model}
Evolved stars such as HD~89345 often show correlated flux variations on the timescales of minutes to hours due to the combination of granulation and pulsation. If unaccounted for, the correlated noise will bias the estimation of transit parameters \citep{carter2009}. To model the correlated flux variation, we employed a Gaussian Process regression which is often used to model stellar variability seen in radial velocity variation of planet host stars \citep[e.g.][]{haywood2014, dai2017}. Here, we adopted a square exponential kernel similar to \citet{grunblatt2016}:
\begin{equation}
\label{covar}
C_{i,j} = h^2 \exp{\left[-\frac{(t_i-t_j)^2}{2\tau^2}\right]}+\sigma^2\delta_{i,j}
\end{equation}
where $C_{i,j}$ are the elements of the covariance matrix, $\delta_{i,j}$ is the Kronecker delta function,
$h$ is the amplitude of the covariance, $t_i$ is the time of $i$th flux observation, $\tau$ is the correlation timescale and $\sigma$ is the white noise component. The set of parameters $h$, $\tau$ and $\sigma$ are known as the hyperparameters of the kernel.

With the above covariance matrix, our likelihood function takes the following form:
\begin{equation}
\label{likelihood}
\log{\mathcal{L}} =  -\frac{N}{2}\log{2\pi}-\frac{1}{2}\log{|\bf{C}|}-\frac{1}{2}\bf{r}^{\text{T}}\bf{C} ^{-\text{1}} \bf{r}
\end{equation}
where $\mathcal{L}$ is the likelihood, $N$ is the number of flux measurements, $\bf{C}$ is the covariance matrix, and $\bf{r}$ is the residual vector i.e. the observed flux variation minus the transit model from {\tt Batman} as described in the previous section.

\subsection{Radial Velocity Model}
\label{sec:rvmodel}
The final component of our joint analysis is a Keplerian model for the measured radial velocity variations of the host star. For a circular orbit, the three parameters of the Keplerian models are the RV semi-amplitude $K$, the orbital period $P_{\text{orb}}$ and time of conjunction $t_{\text{c}}$. We also experimented with an eccentric orbit, which introduces two additional parameters: the eccentricity $e$ and the argument of periastron $\omega$. For unbiased sampling, we transformed these parameters to
$\sqrt{e}$cos$\omega$ and $\sqrt{e}$sin$\omega$ \citep{lucy1971,anderson2011}. For each of the spectrographs we used, we included a systematic offset $\gamma$ and a jitter $\sigma_{\rm jit}$ parameter which subsumes any additional instrumental and stellar noise.

The likelihood function for the radial velocity measurement takes the following form:
\begin{equation}\label{rv_likelihood}
\mathcal{L}=  \prod_{i}\left({\frac{1}{\sqrt{2 \pi (\sigma_i^2 + \sigma_{\text{jit}}(t_i)^2)}}}  \exp \left[ - \frac{[RV(t_i) - \mathcal{M}(t_i)-\gamma(t_i)]^2}{2 (\sigma_i^2 + \sigma_{\text{jit}}(t_i)^2)} \right] \right),
\end{equation}
where $RV(t_i)$ is the measured radial velocity at time $t_i$; $\mathcal{M}(t_i)$ is the Keplerian model at time $t_i$; $\sigma_{i}$ is the internal measurement uncertainty; $\sigma_{\text{jit}}(t_i)$ 
and $\gamma(t_i)$ are the jitter and offset parameters depending on which instrument was used to obtain the measurement $RV(t_i)$.

To avoid confusion with the Rossiter-McLaughlin effect, we exclude RV points taken within 8 hour window around the predicted mid-transit time from this analysis. These data points are modeled separately (see Section~\ref{sec:rm}).

\begin{figure*}
	\includegraphics[width=\columnwidth]{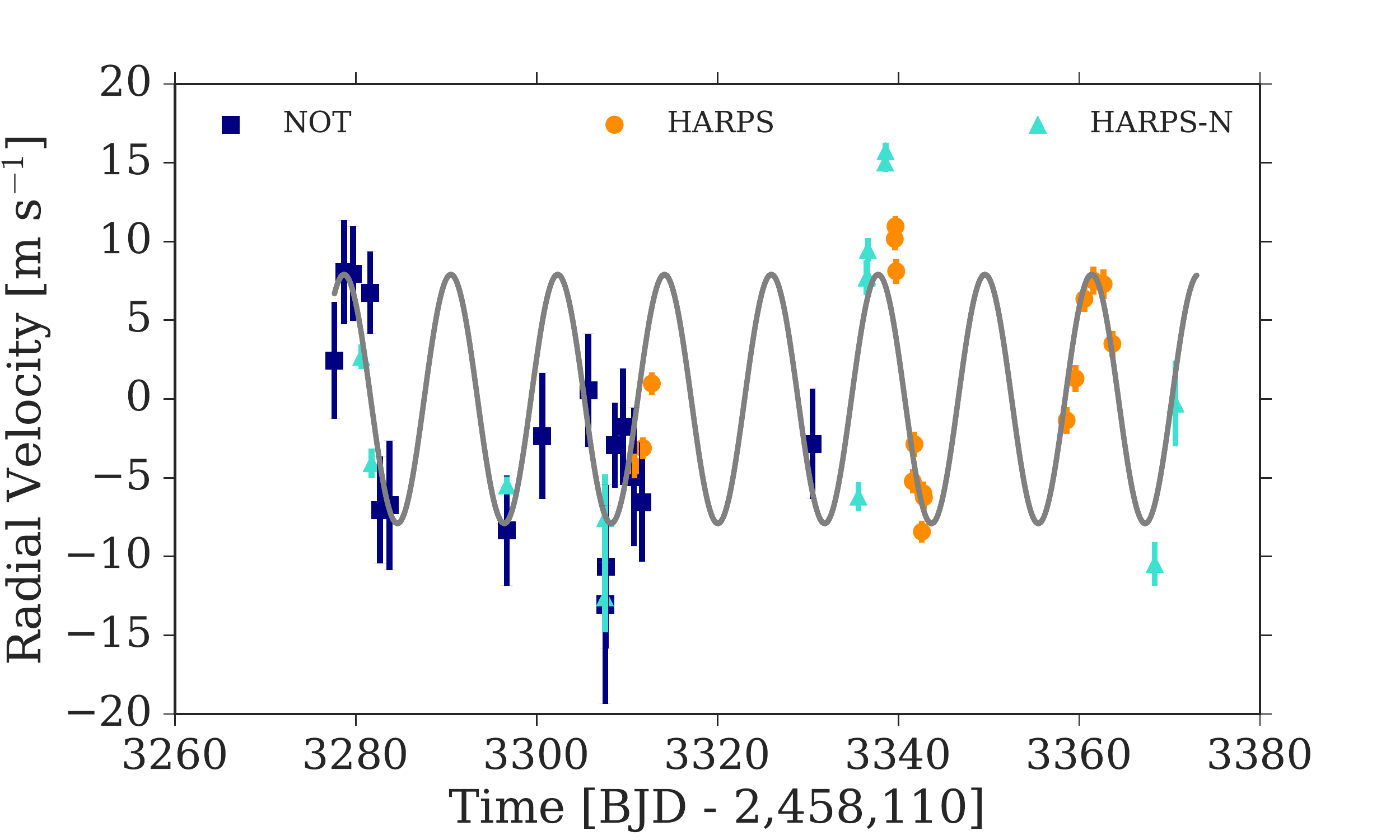}
	\includegraphics[width=\columnwidth]{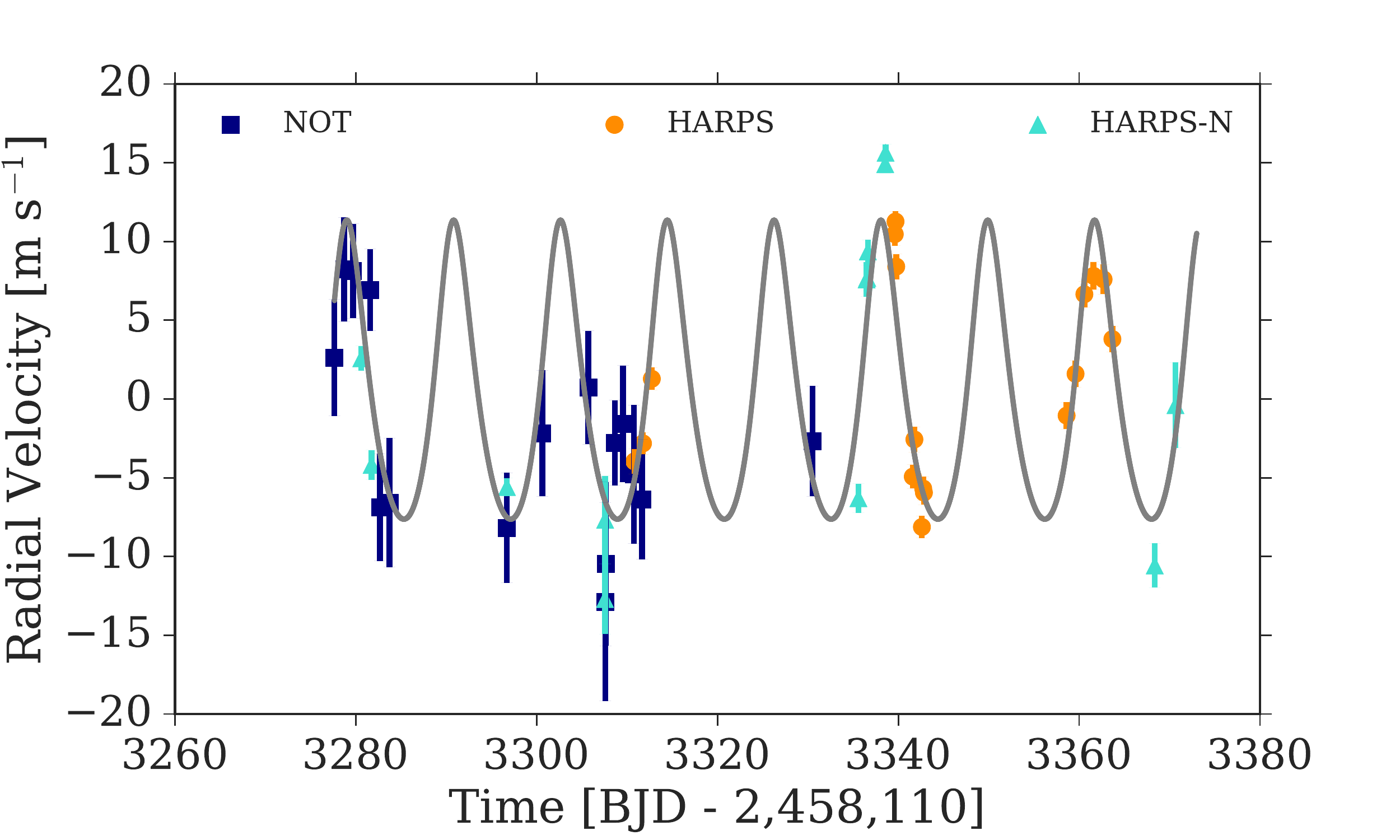}\\
	\includegraphics[width=\columnwidth]{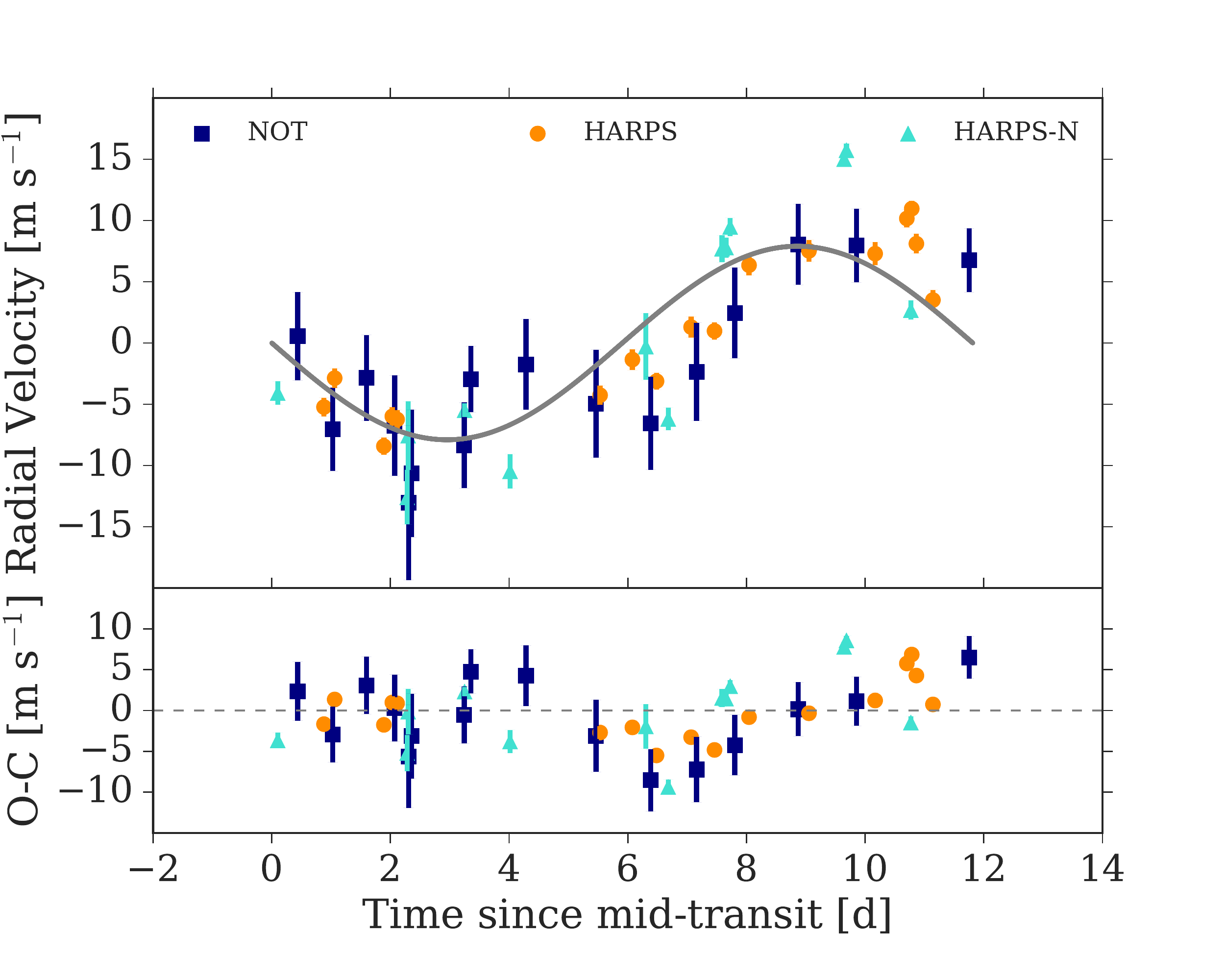}
	\includegraphics[width=\columnwidth]{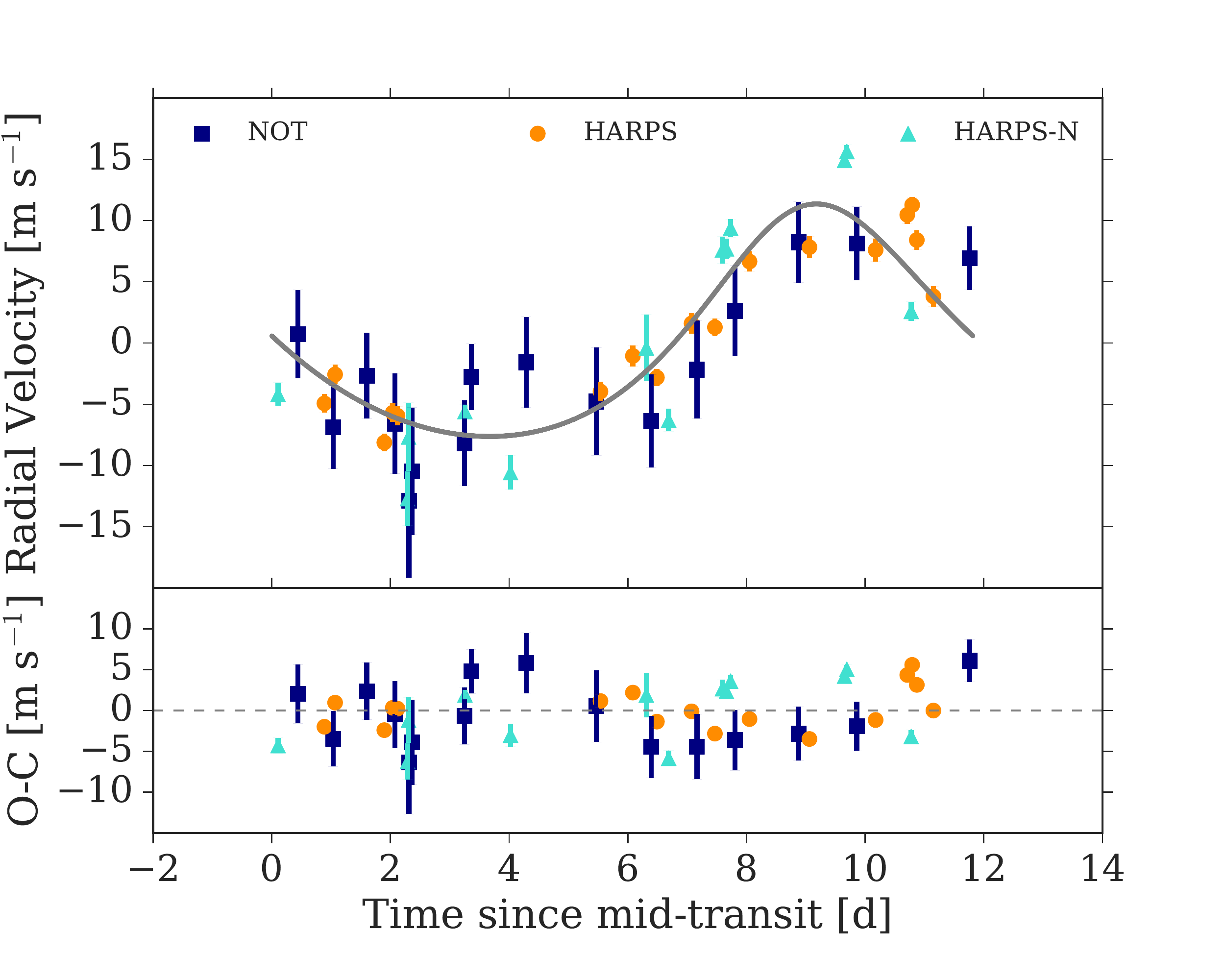}
	\caption{Radial velocity measurements from FIES, HARPS, and HARPS-N are indicated in different colors and symbols. RV points within 8 hours of the transit window are excluded. The top plots show the observations as a function of time. The bottom plots show the observation as a function of phase and include the residuals (observed minus calculated, O-C). In the plots on the left, the best circular model is plotted. In the plots on the right, the best eccentric model is plotted. The observations are provided in Table~\ref{tab:rvdata} and the best values for the models are given in Table~\ref{tab:parameters}.}
	\label{fig:rv}
\end{figure*}

\subsection{Joint analysis}
\label{sec:joint}

To summarize, the free parameters in our joint analysis include the orbital period $P_{\text{orb}}$, the mid-transit time $t_{\text{c}}$, the planet-to-star radius ratio $R_{\text{p}}/R_\star$; the scaled orbital distance $a/R_\star$; the impact parameter $b\equiv a\cos i/R_\star$; the limb-darkening profile $u_1$ and $u_2$; the orbital eccentricity parameters $\sqrt{e}$cos$\omega$ and $\sqrt{e}$sin$\omega$; the amplitude of the covariance $h$; the correlation timescale  $\tau$; the white noise component of the light curve $\sigma$; the RV semi-amplitude $K$; the systematic offset and jitter for each spectrograph $\gamma$, $\sigma_{\rm jit}$. We sampled all the scale parameters ($P_{\text{orb}}$, $R_{\text{p}}/R_\star$, $a/R_\star$, $h$, $\tau$,  $\sigma$, $\sigma_{\rm jit}$) uniformly in $\log$ space, which effectively imposes the Jeffreys prior. We included a prior on the mean stellar density inferred from the asteroseismic analysis $\rho_\star = 0.343 \pm 0.002$~g~cm$^{-3}$ using Equation~30 of \citet{winn2010}. We imposed Gaussian priors on the limb-darkening coefficients $u_1$ and $u_2$ using the median values from EXOFAST\footnote{\url{astroutils.astronomy.ohio-state.edu/exofast/limbdark.shtml}.} \citep{eastman2013} and widths of 0.2. We imposed a uniform prior on the other parameters.

Our final likelihood function is the simple addition of Equation~\ref{likelihood} and the natural logarithm of Equation~\ref{rv_likelihood}. We first located the best-fit solution using the Levenberg-Marquardt algorithm implemented in the {\tt Python} package {\tt lmfit}. We show the best-fit transits, including Gaussian processes, in Figure~\ref{fig:transit}, the best-fit folded transit in Figure~\ref{fig:combotransit}, and the best radial velocity model for both the circular and the eccentric case in Figure~\ref{fig:rv}. To sample the posterior distribution of various parameters, we ran an MCMC analysis with {\tt emcee} \citep{foremanmackey2013}.  We initialized 128 walkers in the vicinity of the best-fit solution. We ran the walkers for 5000 steps and discarded the first 1000 steps as the burn-in phase. We report all parameters in Table~\ref{tab:parameters} using the 16, 50, and 84\,\% percentile cumulative posterior distribution.

\subsection{Orbital eccentricity}
\label{sec:eccentricity}

We find a best-fit orbital eccentricity of $0.203 \pm 0.031$. However, a perfectly circular orbit also provides a reasonable fit to the data, despite the smaller number of parameters. We used the Bayesian Information Criterion (BIC) to check on whether adding the additional two degrees of freedom for an eccentric orbit is justified. We have $5300$ flux observations and 46 RV measurements. The circular model contains 15 parameters, while the eccentric model contains 17. We find a difference in BIC values of 19 between the eccentric fit and the circular fit, favoring the eccentric solution.

When the mean stellar density is known from external observations, the transit duration contains information about the orbital eccentricity \citep[e.g.][]{ford2008}. We investigated the resulting constraint on the eccentricity by fitting the transit data alone (not taking into account the RV observations). Following the procedure described by \cite{vaneylen2015}, we found $e = 0.10^{+0.07}_{-0.10}$, with an uncertainty that is strongly correlated with that of the impact parameter.  Lower impact parameters correspond to higher eccentricity. Alternatively, this measurement shows that the stellar density that can be derived from the transit photometry is consistent with that of the asteroseismic analysis, for near-circular orbits. We note that this solution did not make use of the Gaussian processes described above, but nevertheless resulted in consistent planetary parameters. This solution is consistent with both a circular orbit and with the eccentric fit solution to the combined transit and RV data, at the 95\% confidence level.

In Table~\ref{tab:parameters}, we list all parameters for both the circular and the eccentric solution. However, as the eccentric solution is favored by the data, we adopt these values in the discussion below.

\begin{table*}
  \begin{center}
    \caption{System parameters of HD~89345 (K2-234; EPIC~248777106).\label{tab:parameters}}
    \smallskip
    \begin{tabular}{l c c }
      \hline\hline
      \noalign{\smallskip}
      \multicolumn{3}{c}{Basic properties} \\
      \noalign{\smallskip}
      \hline
      \noalign{\smallskip}
      2MASS ID	 				& \multicolumn{2}{c}{10184106+1007445} \\
      Right Ascension				& \multicolumn{2}{c}{10 18 41.06}  	\\ 
      Declination				& \multicolumn{2}{c}{+10 07 44.50}  	\\
      Magnitude (\textit{Kepler})		& \multicolumn{2}{c}{9.204}  		\\
      Magnitude ($V$)				& \multicolumn{2}{c}{9.30}		\\
      Magnitude ($J$)				& \multicolumn{2}{c}{7.98}		\\
      \hline
      \noalign{\smallskip}
      \multicolumn{3}{c}{Adopted stellar parameters} \\
      \noalign{\smallskip}
      \hline
      \noalign{\smallskip}
      Effective Temperature, $T_{\rm_{eff}}$ (K) 			& \multicolumn{2}{c}{$5499 \pm 73$ }	\\
      Stellar luminosity, $L (\mathrm{L}_\odot$)			& \multicolumn{2}{c}{$2.27^{+0.21}_{-0.14}$ }\\ [3pt]
      Surface gravity, $\log g$ (cgs)            			& \multicolumn{2}{c}{$4.044^{+0.006}_{-0.004}$}	\\ [3pt]
      Metallicity, [Fe/H]                        			& \multicolumn{2}{c}{$0.45 \pm 0.04$ }	\\
      Projected rotation speed, $v \sin i$ (km\,s$^{-1}$) 	& \multicolumn{2}{c}{$2.60 \pm 0.50$}  \\
      Stellar Mass,   $M_{\star} $ ($M_{\odot}$)			& \multicolumn{2}{c}{$1.120^{+0.040}_{-0.010}$}\\[3pt]
      Stellar Radius, $R_{\star} $ ($R_{\odot}$)	 		& \multicolumn{2}{c}{$1.657^{+0.02}_{-0.004}$} \\[3pt]
      Stellar Density, $\rho_\star$ (g cm$^{-3}$)	 		& \multicolumn{2}{c}{$0.343 \pm 0.002$}	 \\
      Age (Gyr)								& \multicolumn{2}{c}{$9.4^{+0.4}_{-1.3}$}	\\
      \hline
       \noalign{\smallskip}
      Parameters from RV and transit fit				& Circular fit			& Eccentric fit (adopted)\\
       \noalign{\smallskip}
       \hline
       \noalign{\smallskip}
      Orbital Period, $P$ (days) 					&  $11.81433 \pm 0.00096$ 	& $11.81399 \pm 0.00086$ \\
      Time of conjunction, $t_{\rm c}$ (BJD$-2454833$) 			&  $3080.80325 \pm 0.00066$ 	& $3080.80316 \pm 0.00062$\\ 
      Orbital eccentricity, $e$ 					& 0 (fixed)		& $0.203 \pm 0.031$\\ 
      Argument of pericenter, $\omega$ ($^{\circ}$) 			& - 				& $-14.9 \pm 20$\\
      Stellar radial velocity amplitude, $K_\star$ (m\,s$^{-1}$)      	& $7.9\pm 1.0$ 			& $9.49\pm 0.84$ \\ 
      Scaled semi-major axis, $a/R_\star$ 				&  $13.628 \pm 0.026$ 		& $ 13.625 \pm 0.027$ \\ 
      Fractional Planetary Radius, $R_{\rm p}/R_\star$ 			&  $0.03840 \pm 0.00025$ 	& $0.03779 \pm 0.00062$\\
      Impact parameter, $b$ 						&  $0.5818 \pm 0.0084$ 		& $0.489 \pm 0.064$\\
      Limb darkening parameter, $u_1$					& $0.47\pm 0.10$ 		& $0.48\pm 0.10$\\\
      Limb darkening parameter, $u_2$ 					& $0.17\pm 0.14$ 		& $0.16\pm 0.13$\\
      \hline      
      Flux white noise $\sigma$ 						&$0.000134\pm 0.000023$ & $0.000134\pm 0.000024$\\
      Covariance amplitude $h$ 							&  $0.0000839 \pm 0.0000058$ & $0.0000836 \pm 0.0000052$\\
      Covariance timescale $\tau$ (days) 					&  $0.00472 \pm 0.00080$ & $0.00485 \pm 0.00079$\\
      Stellar jitter term FIES, $\sigma_{\rm FIES}$ (m\,s$^{-1}$) 		&$<3.2$ 		& $<2.5$\\
      Stellar jitter term HARPS-N, $\sigma_{\rm HARPS-N}$ (m\,s$^{-1}$) 	&  $<6.8$ 		& $<6.3$\\
      Stellar jitter term HARPS, $\sigma_{\rm HARPS}$ (m\,s$^{-1}$) 		& $<3.2$ 		& $<2.8$\\
      Systemic velocity FIES,  	$\gamma_{\rm FIES}$ (m\,s$^{-1}$)	   	& $-2.45\pm 0.91$ 		& $-2.62\pm 0.97$  \\
      Systemic velocity HARPS-N,  $\gamma_{\rm HARPS-N}$ (m\,s$^{-1}$)    	& $2347.4\pm 1.4$		& $2347.5\pm 1.0$\\
      Systemic velocity HARPS, 	$\gamma_{\rm HARPS}$ (m\,s$^{-1}$) 		& $2354.5\pm 0.9$ 	& $2354.2\pm 0.43$\\
	\hline
      \noalign{\smallskip}
      Derived Parameters						& Circular fit			& Eccentric fit (adopted) \\
     \noalign{\smallskip}
      \hline
      \noalign{\smallskip}
      Planetary Mass,   $M_{\rm p} $ ($M_{\oplus}$)			& $30.4 \pm 3.9$		& $35.7 \pm 3.3$ \\ 
      Planetary Radius, $R_{\rm p} $ ($R_{\oplus}$)			& $6.967 \pm 0.096$		& $6.86 \pm 0.14$ \\
      Planetary Density, $\rho_{\rm p}$ (g\,cm$^{-3}$) 			& $0.494 \pm 0.067$		& $0.609 \pm 0.067$ \\
      Semi-major axis, $a$ (AU)	 					& $0.1050 \pm 0.0013$		& $0.1050 \pm 0.0013$ \\
      Equilibrium temperature, $T_\mathrm{eq}$ (K)			& $1053 \pm 14$			& $1053 \pm 14$ \\
    \end{tabular}
  \end{center}
\end{table*}

\section{Discussion}
\label{sec:discussion}

\subsection{Stellar properties}
\label{sec:discussion_star}

\begin{figure}
	\includegraphics[width=\columnwidth]{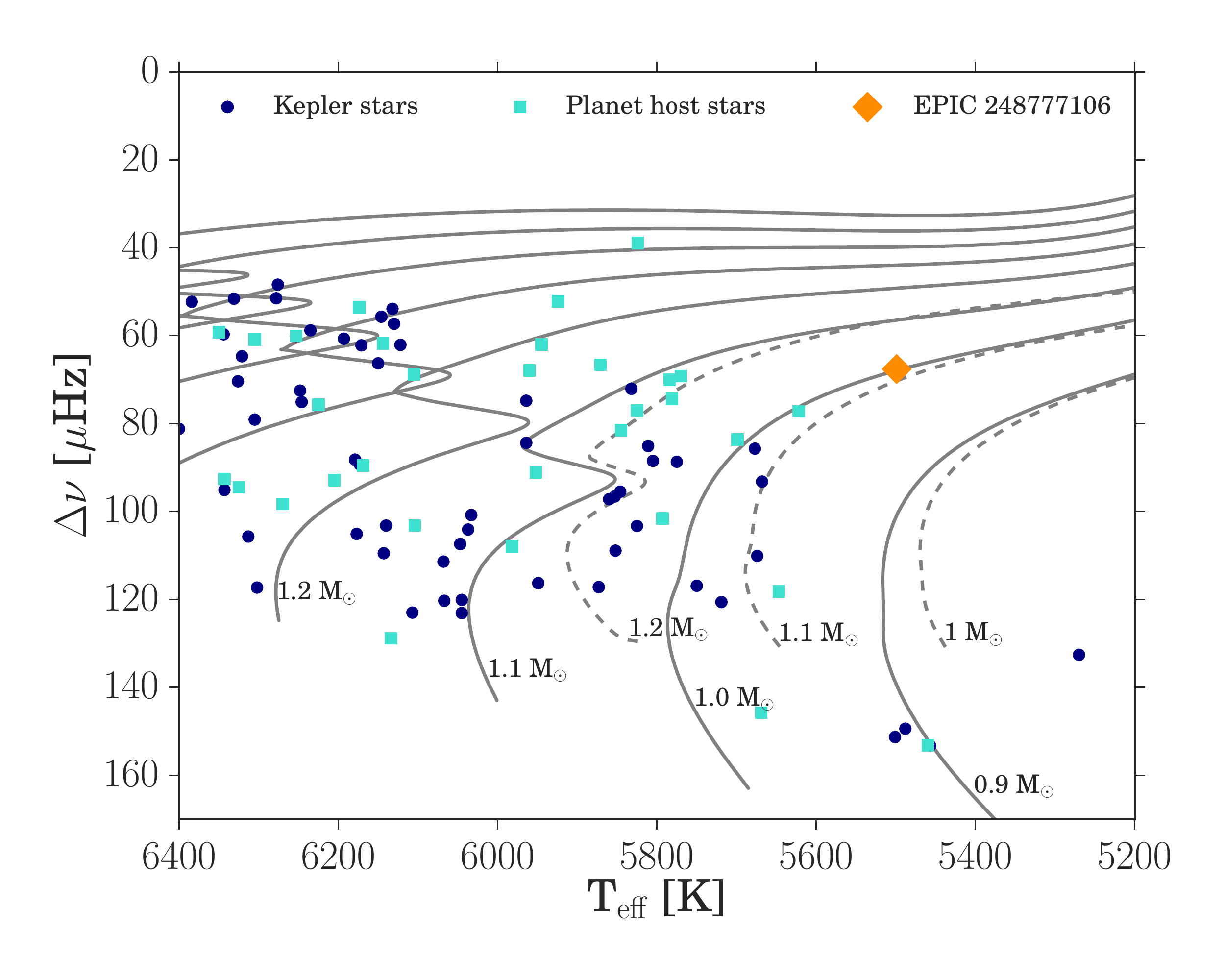}
	\caption{
Modified HR diagram, which depicts the large frequency separation and the effective temperature. In light blue squares, we show solar-like oscillating stars for which the individual frequencies were modeled by \protect\cite{lund2017}. The dark blue circles are the planet-host stars, taken from \protect\cite{davies2016} with a detailed modeling performed by \protect\cite{silvaaguirre2015}. The orange square shows the star analyzed in this work.
Evolution tracks (using the ASTEC models) are shown for a range of masses at solar composition ($Z_\odot$\,=\,0.0246) in grey solid lines and for Fe/H = 0.45~dex (GARSTEC models) in dashed grey lines.}
	\label{fig:hrdiagram}
\end{figure}

HD~89345 is at an interesting phase of its evolution.  The star has just evolved off the main sequence, as can be seen in the Hertzsprung-Russell diagram (see Figure~\ref{fig:hrdiagram}). From the best fit model, it appears to be at the edge of the turn-off point, being a hydrogen shell-burning star with a non-degenerate helium core of 0.06 stellar masses. This explains why no mixed modes were detected in the observed frequency range.
In most previous cases of solar-like oscillators for which the individual frequencies were studied using data from \textit{CoRoT} \citep{baglin2006} or \textit{Kepler} \citep{borucki2010}, the star was either found to be firmly on the main sequence, or firmly on the subgiant branch \citep[e.g.][]{mathur2012,silvaaguirre2015,creevey2017}. Figure~\ref{fig:hrdiagram} shows the stars with asteroseismic analysis of individual oscillation frequencies, for planet-host stars and stars not known to have planets, from the \textit{Kepler} mission. We can see that our target is in a sparsely populated region of this diagram.

Previously, several asteroseismic studies have investigated evolved planet hosting stars, such as subgiant and giant stars, with \textit{Kepler} \citep[e.g.][]{huber2013,huber2013obliquity,silvaaguirre2015,davies2016}, as well as with {K2} \citep[e.g.][]{grunblatt2016,north2017}. The system investigated here is less evolved, and has only just left the main sequence (see Figure~\ref{fig:hrdiagram}). As a result, the oscillation frequencies cannot be detected with the standard long-cadence (30 minute integration) {K2} observations. Here, the availability of short-cadence observations enabled the asteroseismic measurement.

The depth of the convective zone is 32\% of the stellar radius, and the depth of the helium second ionization zone is 3\% of the stellar radius. These values are obtained as the best-fit parameters from the modeling, as $p$ mode oscillations of subgiant stars are very sensitive to the location of these layers \citep[see e.g.][]{grundahl2017}. Both zones are a bit deeper in this star than they are in the Sun. Locating the position of the base of the convective zone is interesting in order to better understand the mechanism of the stellar dynamo, while the helium second ionization zone provides insights in the process of chemical enrichment in stars.

\subsection{Planet properties}
\label{sec:discussion_planet}

HD~89345b is a sub-Saturn planet, with a radius of $6.86 \pm 0.14~R_\oplus$. In the solar system, no planets exist with a size between Uranus ($4~R_\oplus$) and Saturn ($9.45~R_\oplus$). Sub-Saturn planets span a wide range of masses, spanning from $6$ to $60~M_\oplus$, independent of their size \citep{petigura2017}. Although similar to Jovian planets in that they have a large envelope of hydrogen and helium gas, sub-Saturns have much lower masses.  This suggests that sub-Saturns did not undergo runaway gas accretion.   Alternative scenarios have been proposed, such as accretion within a depleted gas disk \citep{lee2015}.

\begin{figure*}
	\includegraphics[width=\columnwidth]{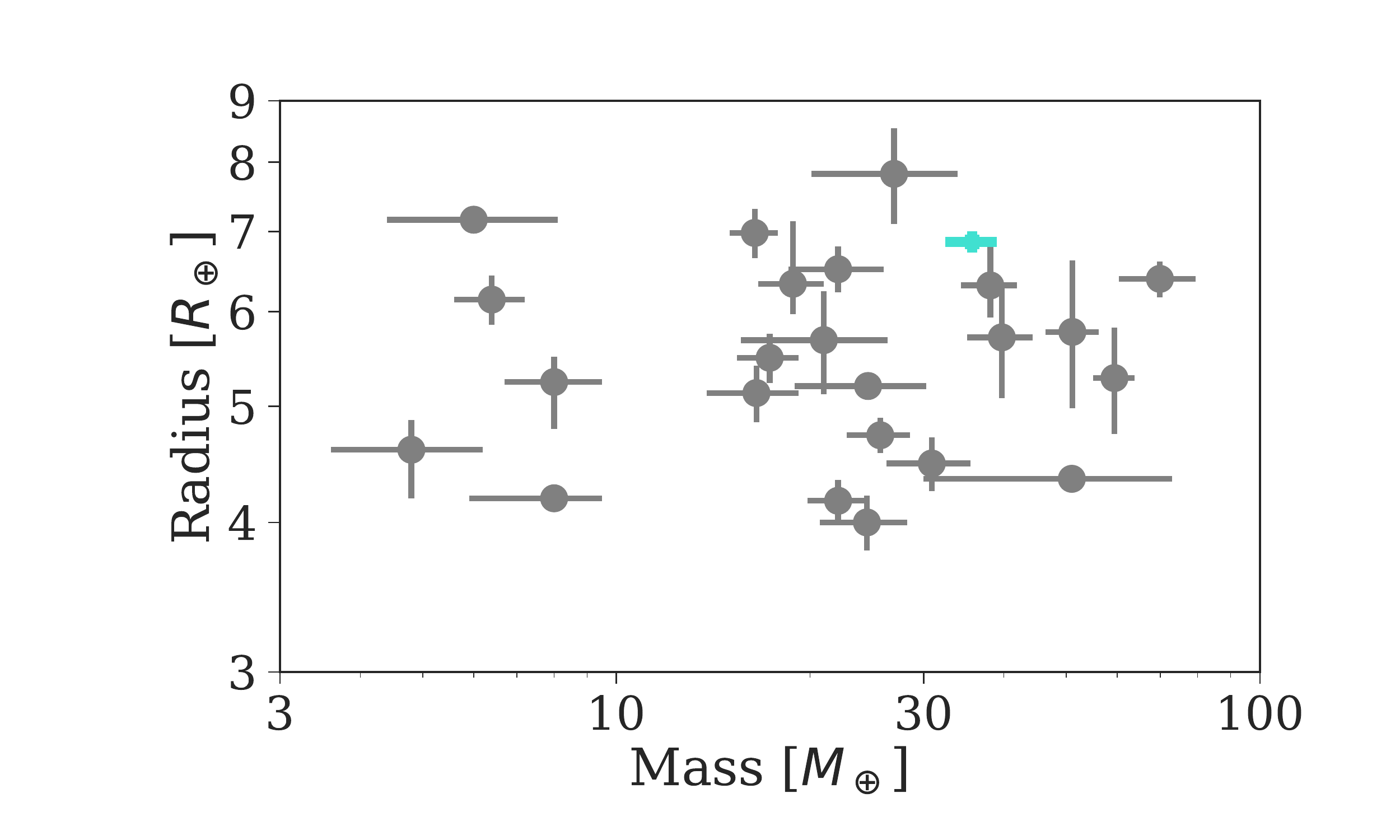}
	\includegraphics[width=\columnwidth]{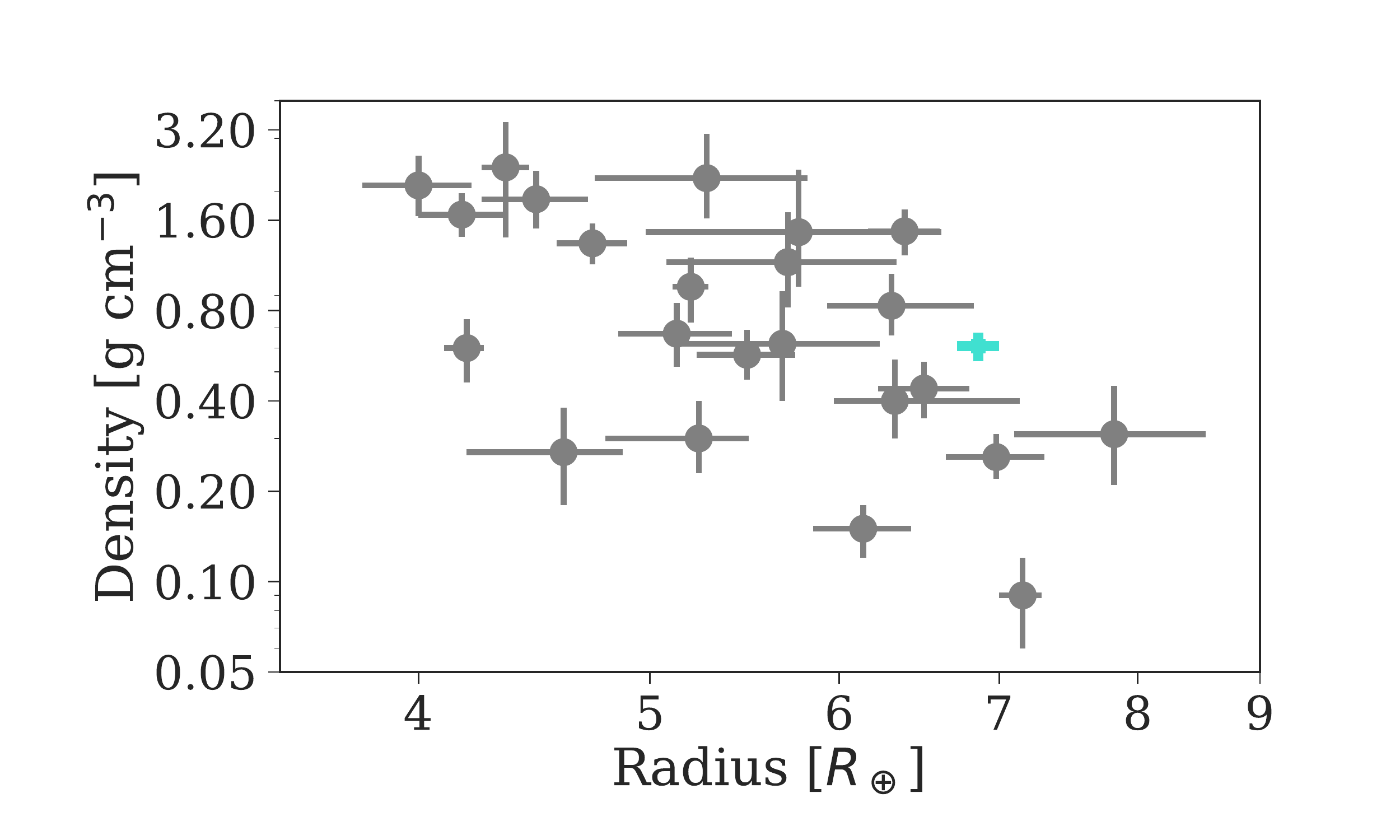}
	\includegraphics[width=\columnwidth]{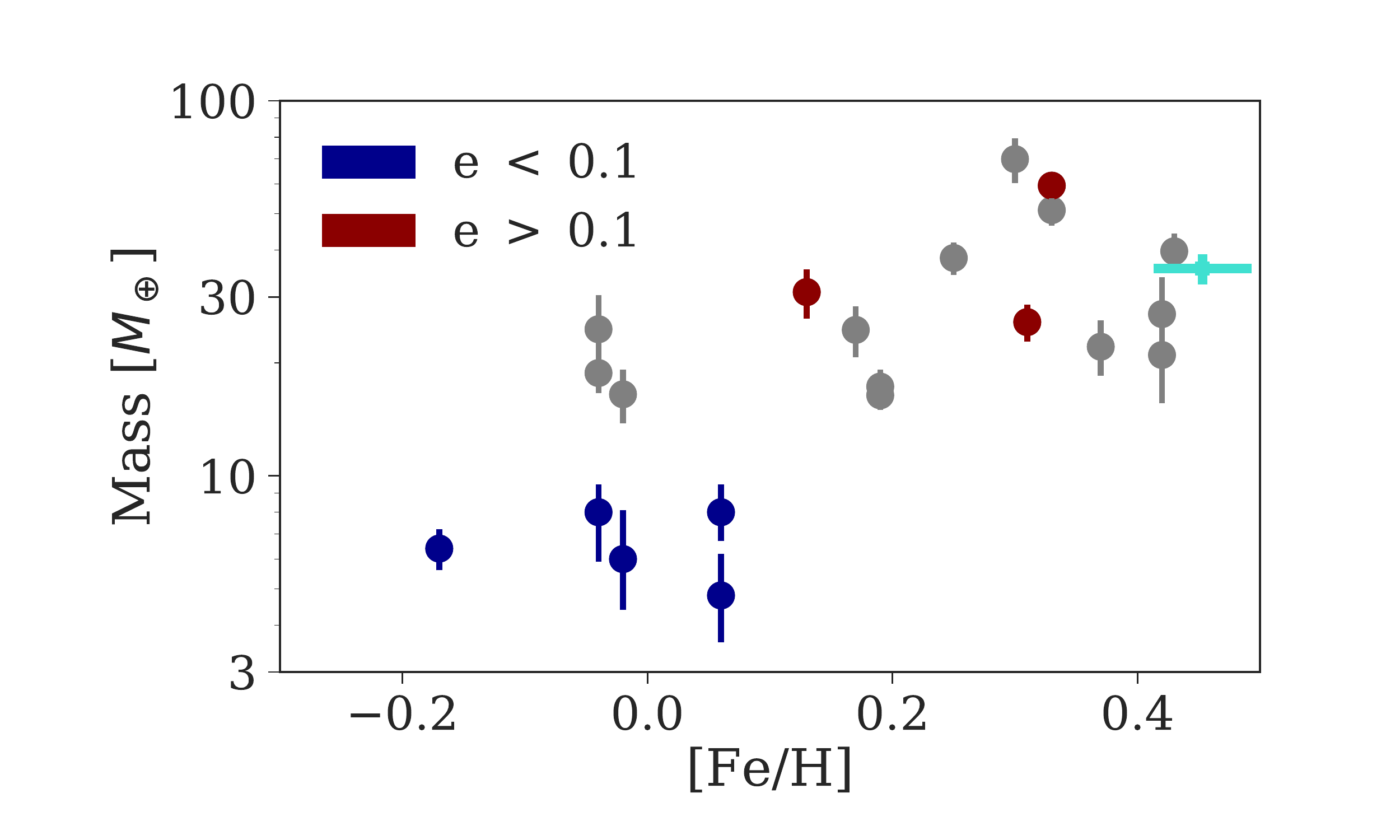}
	\includegraphics[width=\columnwidth]{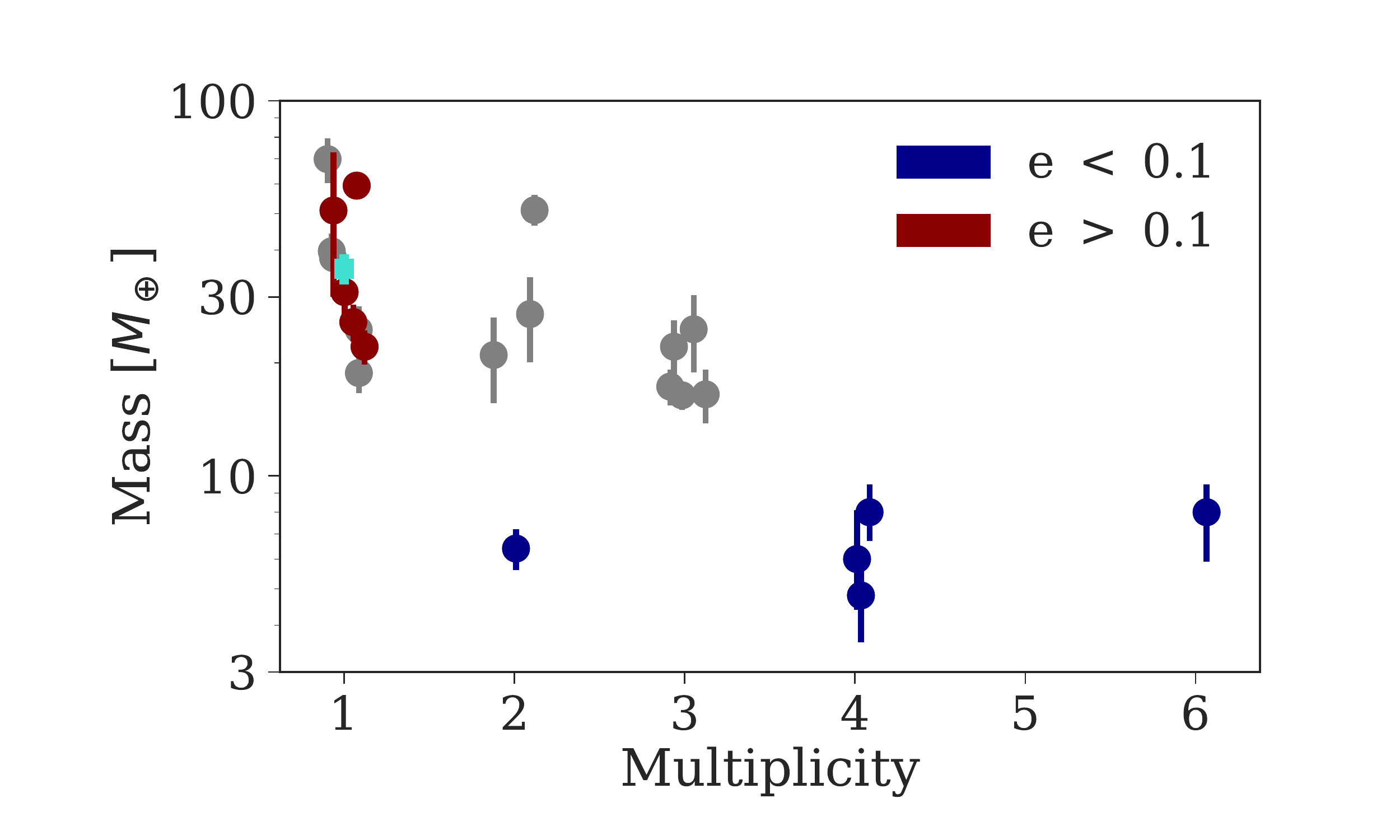}
	\caption{The properties of sub-Saturns as listed in \protect\cite{petigura2017} are listed in grey. In the bottom plot, we use red and blue symbols when the eccentricity is clearly established \protect\citep[again following][]{petigura2017}, for eccentric and circular orbits respectively. HD~89345b is shown in a light blue square. As a relatively high-mass sub-Saturn planet, HD~89345b fits the pattern as a single detected planet with a significant eccentricity orbiting a metal-rich star. For the multiplicity, the values are slightly offset for clarity.}
	\label{fig:massradius}
\end{figure*}

HD~89345b joins a list of 24 sub-Saturns with a mean density measured to better than 50\% \citep[see][Table 7]{petigura2017}. In these systems, \cite{petigura2017} find that higher-mass planets are associated with a higher stellar metallicity, a low planet multiplicity, and a non-zero orbital eccentricity. HD~89345b has a relatively high mass, orbits a star with a relatively high metallicity, is the only detected planet in the system, and appears to have an eccentric orbit. It therefore fits all of these expectations,
as shown in Figure~\ref{fig:massradius}.

Here, we have adopted the planet's eccentric orbital solution. We estimate the timescale of circularization following \cite{goldreich1966} and using a modified tidal quality factor of $Q' = 10^5$ as suggested by \cite{petigura2017}, and find a circularization timescale of 18 Gyr, suggesting that if the orbit was eccentric early in its formation, it could still be eccentric today. 
However, recent high-precision astrometric data obtained with the CASSINI space mission suggest a stronger value for the current tidal dissipation in Saturn, with a modified tidal quality factor $Q^{'}\approx 9434$ \citep{lainey2017}. Assuming such a value, which can be explained by different ab-initio models of tidal dissipation both in the potential rocky/icy core of the planet \citep{remus2012,lainey2017} or in its fluid envelope \citep{ogilvie2004,guenel2014,fuller2016}, the circularisation timescale will be shorter, i.e.\ 1.69~Gyr, a value that is also compatible with the age of the host star. Therefore, the apparent eccentric orbit suggests a weaker dissipation in warm Saturns than in Saturn, which is similar to the weaker dissipation in hot Jupiters than in Jupiter, as has been previously suggested \citep[][and references therein]{ogilvie2014}.

The flux of radiation that the planet receives from the star is roughly 150 times the flux the Earth receives from the Sun.  Thus the planet is heavily irradiated, but not quite at the level at which evidence of photo-evaporation is seen \citep{fulton2017,vaneylen2017}.

\subsection{Future work}

We investigated the rotational splitting of the stellar oscillations, which have the potential to reveal the stellar inclination angle. However, the posterior distribution of this analysis is consistent with a wide range of stellar inclination angles. Similarly, Rossiter-McLaughlin observations cannot reliably constrain the stellar obliquity. Future such measurements, although challenging for shallow transits, may lead to a clearer detection of the Rossiter-McLaughlin effect, owing to the brightness of the host star. The medium-level impact parameter further facilitates such studies.

Due to its low density, HD~89345b may be a target for atmospheric characterization. However, given the large stellar radius, the expected transmission signal per scale height (H) of the planetary atmosphere, assuming an H$_2$/He dominated atmosphere with $\mu=2.3$, is only 48~parts per million (ppm). Under the same assumption, and also assuming that its atmosphere exhibits pure Rayleigh scattering, the transit depth difference between g' and z' bands would be about 140 ppm \citep[see e.g.][for details]{madhusudhan2016}. If the mean molecular weight were closer to that of Neptune rather than Jupiter, the transmission signal would be even smaller. Given these numbers, atmospheric characterization would likely be out of reach for most instruments, except perhaps for the James Webb Space Telescope \citep{gardner2006}.

Asteroseismology of planet host stars has been a fruitful endeavor with the \textit{Kepler} mission, but has so far been limited to evolved stars for {K2}. This is the least evolved planet host star for which asteroseismology has been possible with only 80 days of {K2} observations. 

The detection of individual stellar oscillation modes, and even moderate constraints on the rotational splittings, with 80 days of photometry, is encouraging for asteroseismic detection with the upcoming \textit{TESS} mission \citep{ricker2014} which will provide one month of observations for most bright stars in the sky, as well as longer photometric time series for certain regions of the sky.

\section*{Acknowledgements}

We gratefully acknowledge many helpful suggestions by the anonymous referee. Based on observations made with a) the Nordic Optical
Telescope, operated by the Nordic Optical Telescope Scientific
Association at the Observatorio del Roque de los Muchachos; b) the
ESO-3.6m telescope at La Silla Observatory under programme ID
0100.C-0808; c) the Italian Telescopio Nazionale Galileo operated on
the island of La Palma by the Fundaci\'on Galileo Galilei of the
Istituto Nazionale di Astrofisica. 
NESSI was funded by the NASA Exoplanet Exploration Program and the NASA Ames Research Center. NESSI was built at the Ames Research Center by Steve B. Howell, Nic Scott, Elliott P. Horch, and Emmett Quigley. 
This project has received funding
from the European Union's Horizon 2020 research and innovation
programme under grant agreement No 730890. This material reflects
only the authors views and the Commission is not liable for any use
that may be made of the information contained therein.
DG gratefully acknowledges the financial support of the \emph{Programma Giovani Ricercatori -- Rita Levi Montalcini -- Rientro dei Cervelli (2012)} awarded by the Italian Ministry of Education, Universities and Research (MIUR). 
SaM would like to acknowledge support from the Ramon y Cajal fellowship number RYC-2015-17697.
AJ, MH, and SA acknowledge support
by the Danish Council for Independent Research, through a DFF Sapere
Aude Starting Grant nr.\ 4181-00487B.
SzCs, APH, MP, and HR acknowledge the support of the DFG priority program
SPP 1992 "Exploring the Diversity of Extrasolar Planets (grants HA
3279/12-1, PA 525/18-1, PA5 25/19-1 and PA525/20-1, RA 714/14-1)
HD, CR, and FPH acknowledge the financial support from MINECO under grants ESP2015-65712-C5-4-R and AYA2016-76378-P. This paper has made use of the IAC Supercomputing facility HTCondor (\url{http://research.cs.wisc.edu/htcondor/}), partly financed by the Ministry of Economy and Competitiveness with FEDER funds, code IACA13-3E-2493. 
MF and CMP gratefully acknowledge the support of the 
Swedish National Space Board.
RAG and StM thanks the support of the CNES PLATO grant.
PGB is a postdoctoral fellow in the MINECO-programme 'Juan de la Cierva Incorporacion' (IJCI-2015-26034).
StM acknowledges support from ERC through SPIRE grant (647383) and from ISSI through the ENCELADE 2.0 team.
VSA acknowledges support from VILLUM FONDEN (research grant 10118).
MNL acknowledges support from the ESA-PRODEX programme. 
Funding for the Stellar Astrophysics Centre is provided by The Danish National Research Foundation (Grant agreement no.: DNRF106)
This work has made use of data from the European Space Agency (ESA) mission
{\it Gaia} (\url{https://www.cosmos.esa.int/gaia}), processed by the {\it Gaia}
Data Processing and Analysis Consortium (DPAC,
\url{https://www.cosmos.esa.int/web/gaia/dpac/consortium}). Funding for the DPAC
has been provided by national institutions, in particular the institutions
participating in the {\it Gaia} Multilateral Agreement.
This research was made with the use of NASA's Astrophysics Data System and the NASA Exoplanet Archive, which is operated by the California Institute of Technology, under contract with the National Aeronautics and Space Administration under the Exoplanet Exploration Program.




\balance
\bibliographystyle{mnras}
\bibliography{references_asteroseismic,references_k2} 


\appendix
\section{Extra material	}
%

%
\begin{table*}
\scriptsize
\caption{Radial velocity observations (see Section~\ref{sec:observations_spectro} for details).	
\textbf{Notes.}
$^{1}$SNR calculated per pixel at 5500 \AA . $^{2}$ The SNR in the blue part of the spectrum was too low to calculate these $\log_\mathrm{RHK}$ values. \label{tab:rvdata}}
\begin{center}
 \begin{tabular}{ cccccccccc } 
       \hline\hline
      \noalign{\smallskip}
      Time &  RV	& $\sigma_{\mathrm{RV}}$ & Ins.\ & BIS & FWHM & $\log_\mathrm{RHK}$ & 
      $\sigma(\log_\mathrm{RHK})$ & t$_\mathrm{exp}$ & SNR$^{(1)}$ \\
      
      [BJD] &  [km s$^{-1}$]	& [km s$^{-1}$] & & [km s$^{-1}$] & [km s$^{-1}$]& & 
       &  & \\
      \noalign{\smallskip}
      \hline
  2458110.635193 &  0.0000 & 0.0037 & FIES &  0.0004 & 12.6730 &  -  &  -  & 1800.0 &  98.8 \\
  2458111.705789 &  0.0056 & 0.0033 & FIES &  0.0001 & 12.6718 &  -  &  -  & 1800.0 &  71.2 \\
  2458112.686073 &  0.0055 & 0.0030 & FIES &  0.0022 & 12.6721 &  -  &  -  & 1200.0 &  97.0 \\
  2458114.586884 &  0.0043 & 0.0026 & FIES &  0.0040 & 12.6676 &  -  &  -  & 2400.0 &  98.1 \\
  2458115.674859 & -0.0095 & 0.0034 & FIES &  0.0047 & 12.6412 &  -  &  -  & 1800.0 &  68.6 \\
  2458116.716467 & -0.0092 & 0.0041 & FIES &  0.0012 & 12.6529 &  -  &  -  & 1800.0 &  65.4 \\
  2458129.703737 & -0.0108 & 0.0035 & FIES & -0.0021 & 12.6590 &  -  &  -  & 1800.0 &  73.0 \\
  2458133.620112 & -0.0048 & 0.0040 & FIES & -0.0042 & 12.6488 &  -  &  -  & 1800.0 &  70.7 \\
  2458138.710781 & -0.0019 & 0.0036 & FIES & -0.0056 & 12.6181 &  -  &  -  & 1800.0 &  65.2 \\
  2458140.581356 & -0.0155 & 0.0063 & FIES & -0.0045 & 12.6574 &  -  &  -  & 3600.0 &  39.1 \\
  2458140.630614 & -0.0131 & 0.0052 & FIES & -0.0036 & 12.6471 &  -  &  -  & 3600.0 &  44.3 \\
  2458141.633063 & -0.0054 & 0.0027 & FIES &  0.0136 & 12.6659 &  -  &  -  & 1800.0 &  97.1 \\
  2458142.559006 & -0.0042 & 0.0037 & FIES &  0.0000 & 12.6593 &  -  &  -  & 2700.0 &  67.7 \\
  2458143.739530 & -0.0074 & 0.0044 & FIES & -0.0032 & 12.6387 &  -  &  -  & 3600.0 &  59.0 \\
  2458144.662520 & -0.0090 & 0.0038 & FIES & -0.0027 & 12.6575 &  -  &  -  & 1800.0 &  68.8 \\
  2458163.498602 & -0.0053 & 0.0035 & FIES & -0.0016 & 12.6668 &  -  &  -  & 1800.0 &  70.5 \\
  2458143.808287 & 2.3502 & 0.0008 & HARPS &  0.0018 & 7.6348 & -5.1690 & 0.0138 & 1200.0 &  98.4 \\
  2458144.759327 & 2.3514 & 0.0007 & HARPS &  0.0007 & 7.6311 & -5.1684 & 0.0111 & 1200.0 & 109.4 \\
  2458145.736594 & 2.3555 & 0.0007 & HARPS &  0.0021 & 7.6289 & -5.1846 & 0.0116 & 1500.0 & 106.9 \\
  2458172.606520 & 2.3646 & 0.0007 & HARPS &  0.0016 & 7.6303 & -5.1321 & 0.0096 & 1200.0 & 101.7 \\
  2458172.689287 & 2.3654 & 0.0006 & HARPS &  0.0060 & 7.6271 & -5.1589 & 0.0095 & 1200.0 & 116.1 \\
  2458172.767539 & 2.3626 & 0.0008 & HARPS & -0.0033 & 7.6303 & -5.1781 & 0.0153 &  900.0 &  97.2 \\
  2458173.566565 & 2.3594 & 0.0007 & HARPS &  0.0026 & 7.6321 & -5.1514 & 0.0111 & 1200.0 & 100.2 \\
  2458173.580118 & 2.3613 & 0.0008 & HARPS &  0.0004 & 7.6283 & -5.1327 & 0.0104 & 1200.0 &  98.8 \\
  2458173.594505 & 2.3599 & 0.0009 & HARPS &  0.0042 & 7.6364 & -5.1725 & 0.0137 & 1200.0 &  88.3 \\
  2458173.609169 & 2.3590 & 0.0009 & HARPS &  0.0057 & 7.6253 & -5.1520 & 0.0138 & 1200.0 &  85.5 \\
  2458173.623405 & 2.3601 & 0.0008 & HARPS &  0.0062 & 7.6299 & -5.1486 & 0.0113 & 1200.0 &  95.8 \\
  2458173.637236 & 2.3588 & 0.0007 & HARPS &  0.0032 & 7.6329 & -5.1351 & 0.0102 & 1200.0 & 100.5 \\
  2458173.651484 & 2.3609 & 0.0008 & HARPS &  0.0022 & 7.6302 & -5.1681 & 0.0123 & 1200.0 &  93.7 \\
  2458173.666009 & 2.3574 & 0.0008 & HARPS &  0.0041 & 7.6284 & -5.1393 & 0.0124 & 1200.0 &  90.2 \\
  2458173.680106 & 2.3576 & 0.0007 & HARPS &  0.0029 & 7.6354 & -5.1305 & 0.0104 & 1200.0 & 103.4 \\
  2458173.694493 & 2.3546 & 0.0008 & HARPS &  0.0012 & 7.6280 & -5.1445 & 0.0114 & 1200.0 &  98.5 \\
  2458173.708741 & 2.3572 & 0.0007 & HARPS &  0.0021 & 7.6307 & -5.1614 & 0.0109 & 1200.0 & 109.6 \\
  2458173.722838 & 2.3569 & 0.0007 & HARPS &  0.0035 & 7.6322 & -5.1469 & 0.0104 & 1200.0 & 114.9 \\
  2458173.737224 & 2.3584 & 0.0007 & HARPS & -0.0001 & 7.6276 & -5.1525 & 0.0112 & 1200.0 & 105.4 \\
  2458173.751472 & 2.3584 & 0.0007 & HARPS &  0.0020 & 7.6255 & -5.1277 & 0.0103 & 1200.0 & 111.3 \\
  2458173.765569 & 2.3575 & 0.0007 & HARPS &  0.0015 & 7.6329 & -5.1554 & 0.0115 & 1200.0 & 110.8 \\
  2458173.779956 & 2.3572 & 0.0006 & HARPS &  0.0055 & 7.6286 & -5.1393 & 0.0106 & 1200.0 & 119.4 \\
  2458173.794064 & 2.3565 & 0.0006 & HARPS &  0.0024 & 7.6305 & -5.1532 & 0.0111 & 1200.0 & 123.7 \\
  2458173.808312 & 2.3547 & 0.0006 & HARPS &  0.0020 & 7.6343 & -5.1235 & 0.0103 & 1200.0 & 126.9 \\
  2458173.822560 & 2.3564 & 0.0006 & HARPS &  0.0045 & 7.6294 & -5.1662 & 0.0118 & 1200.0 & 125.2 \\
  2458173.836668 & 2.3574 & 0.0007 & HARPS &  0.0026 & 7.6311 & -5.1369 & 0.0123 & 1200.0 & 120.6 \\
  2458173.851043 & 2.3575 & 0.0007 & HARPS &  0.0026 & 7.6329 & -5.1972 & 0.0160 & 1200.0 & 112.4 \\
  2458174.597927 & 2.3492 & 0.0008 & HARPS &  0.0039 & 7.6338 & -5.1507 & 0.0107 & 1200.0 &  98.9 \\
  2458174.777961 & 2.3516 & 0.0008 & HARPS &  0.0013 & 7.6273 & -5.1582 & 0.0140 &  900.0 &  97.1 \\
  2458175.607320 & 2.3460 & 0.0007 & HARPS &  0.0018 & 7.6323 & -5.1414 & 0.0093 & 1200.0 & 104.7 \\
  2458175.749923 & 2.3485 & 0.0008 & HARPS &  0.0023 & 7.6315 & -5.1476 & 0.0135 &  900.0 & 102.5 \\
  2458175.829842 & 2.3482 & 0.0008 & HARPS &  0.0016 & 7.6325 & -5.2029 & 0.0185 & 1080.0 & 103.6 \\
  2458191.610255 & 2.3531 & 0.0009 & HARPS &  0.0060 & 7.6351 & -5.0850 & 0.0107 &  900.0 &  86.8 \\
  2458192.599492 & 2.3558 & 0.0008 & HARPS &  0.0006 & 7.6352 & -5.0962 & 0.0108 & 1050.0 &  87.4 \\
  2458193.575951 & 2.3608 & 0.0008 & HARPS & -0.0022 & 7.6356 & -5.0926 & 0.0099 &  900.0 &  87.7 \\
  2458194.585844 & 2.3620 & 0.0009 & HARPS & -0.0017 & 7.6296 & -5.1048 & 0.0115 &  900.0 &  82.8 \\
  2458195.701147 & 2.3618 & 0.0009 & HARPS &  0.0027 & 7.6295 & -5.0994 & 0.0140 &  900.0 &  80.4 \\
  2458196.675171 & 2.3580 & 0.0008 & HARPS & -0.0010 & 7.6303 & -5.1836 & 0.0142 &  900.0 &  90.4 \\
  2458113.602755 & 2.3502 & 0.0008 & HARPS-N & -0.0029 & 7.6009 & -5.1238 & 0.0108 & 1500.0 &  92.0 \\
  2458114.746684 & 2.3434 & 0.0009 & HARPS-N & -0.0037 & 7.6015 & -5.1046 & 0.0132 &  900.0 &  75.8 \\
  2458129.709781 & 2.3420 & 0.0006 & HARPS-N & -0.0017 & 7.6067 & -5.1241 & 0.0061 & 1800.0 & 124.5 \\
  2458140.557742 & 2.3349 & 0.0022 & HARPS-N & -0.0057 & 7.6049 & -5.0840 & 0.0655 & 1200.0 &  40.2 \\
  2458140.573703 & 2.3399 & 0.0028 & HARPS-N &  0.0026 & 7.6069 & -5.0065 & 0.0626 & 1200.0 &  32.6 \\
  2458169.491793 & 2.3552 & 0.0011 & HARPS-N & -0.0012 & 7.5808 & -5.2133 & 0.0220 & 2400.0 &  68.5 \\
  2458169.559756 & 2.3553 & 0.0008 & HARPS-N & -0.0026 & 7.5893 & -5.1472 & 0.0113 & 2100.0 &  89.4 \\
  2458169.629247 & 2.3569 & 0.0007 & HARPS-N & -0.0033 & 7.5862 & -5.1628 & 0.0099 & 2100.0 &  98.9 \\
  2458171.549472 & 2.3625 & 0.0006 & HARPS-N & -0.0000 & 7.5877 & -5.1531 & 0.0066 & 1500.0 & 123.1 \\
  2458171.588592 & 2.3632 & 0.0005 & HARPS-N & -0.0019 & 7.5879 & -5.1311 & 0.0054 & 1500.0 & 133.5 \\
  2458201.362286 & 2.3370 & 0.0014 & HARPS-N & -0.0017 & 7.5874 & -5.2441 & 0.0365 & 1800.0 &  56.5 \\
  2458203.651234 & 2.3472 & 0.0027 & HARPS-N & -0.0080 & 7.5902 &     -$^{(2)}$       &    -$^{(2)}$       & 1200.0 &  33.9 \\
\end{tabular}
\end{center}
\end{table*}

\begin{table*}
\caption{A list of all detected oscillation frequencies and their uncertainties, derived according to the Bayesian method and using the MAP algorithm (see Section~\ref{sec:analysis_asteroseismic}), together with their radial order and angular degree.}
\label{tab:frequencies}
\centering
\begin{tabular}{c c c c c c c c}
\hline \hline
Order & Degree & Freq.\ (Bayes) [$\mu$Hz] & $\sigma+_\mathrm{freq.\, Bayes}$ [$\mu$Hz] & Freq.\ (MAP) [$\mu$Hz] & $\sigma_{\mathrm{freq., MAP}}$ [$\mu$Hz]\\
\hline
14 &   0 &   		&	& 1036.81	&0.72\\     
14 &   1 &       	&	& 1065.09 	&0.66 \\ 
14 &   2 & 		&	& 1097.22	&0.85\\            
15 &   0 &		&	& 1104.12  	&0.65  \\    
15 &   1 & 		& 	& 1131.34	&0.51 \\
15 &   2 &    1162.99   &0.46 	& 1163.14   	&0.25 \\
16 &   0 &    1168.60   &0.19  	& 1168.64	&0.18  \\
16 &   1 &    1197.36   &0.20  	& 1197.30	&0.17  \\
16 &   2 &    1230.81   &0.28  	& 1230.60   	&0.27\\ 
17 &   0 &    1236.03   &0.96  	& 1235.92   	&0.30\\
17 &   1 &    1264.61   &0.18  	& 1264.83	&0.14  \\
17 &   2 &    1299.19   &0.29  	& 1299.27  	&0.27 \\
18 &   0 &    1303.64   &0.21  	&1303.58   	&0.24 \\
18 &   1 &    1332.55   &0.19  	&1332.56  	&0.17\\
18 &   2 &    1366.61   &0.49  	&1366.70   	&0.38 \\
19 &   0 &    1370.87   &0.28  	&1370.98   	&0.37\\
19 &   1 &    1399.62   &0.30  	&1399.59   	&0.21\\
19 &   2 &    1433.46   &0.51	&1433.56   	&0.33  \\
20 &   0 &    1438.52   &0.50  	&1438.68   	&0.32\\
20 &   1 &    1466.49   &0.36  	&1466.75   	&0.29\\
20 &   2 &    1502.42   &1.1   	&1503.12   	&0.63\\
21 &   0 &    1506.45   &0.26	&1506.34   	&0.39\\
21 &   1 &    1534.18   &0.30 	&1534.50   	&0.46 \\
21 &   2 & 		&	&1569.65   	&1.29\\            
22 &   0 &		&	&1575.00	&1.02\\   
\end{tabular}
\end{table*}

\label{lastpage}
\end{document}